%% file: mesoscale_voids_text_arXiv.tex
\documentclass[10pt,oneside]{article}
\usepackage[sc]{mathpazo}
\usepackage[T1]{fontenc}
\usepackage{latexsym}
\usepackage{tikz}
\usepackage{subfig,sidecap,epstopdf}
\usepackage{color}
\usepackage{graphics}
\usepackage{amsmath}
\usepackage{amssymb}
\usepackage{enumerate}
\usepackage{mathrsfs}
\newtheorem{thm}{Theorem}
\newtheorem{lem}{Lemma}
\newtheorem{coro}{Corollary}

\renewcommand{\theequation}{\thesection.\arabic{equation}}
\numberwithin{equation}{section}

\date{}

\input macroarx

\title{Meso-scale models  and approximate solutions for solids containing clouds of voids}
\author{V.G.  Maz'ya\footnote{Department of Mathematics, Link\"oping University, SE-581 83 Link\"oping, Sweden.}, A.B. Movchan\footnote{Department of Mathematical Sciences, University of Liverpool, Liverpool L69 3BX, U.K.}\,\, and M.J. Nieves\footnote{School of Engineering, Liverpool John Moores University, James Parsons Building, Byrom Street, Liverpool L3 3AF, U.K.}}
\date{}

\begin{document}
\maketitle

\begin{abstract}
For highly perforated domains the paper addresses a novel approach to study mixed boundary value problems for the equations of linear elasticity in the framework of meso-scale approximations. There are no assumptions of periodicity involved in the description of the geometry of the domain. The size of the perforations is small compared to the minimal separation between neighbouring defects and here we discuss a class of problems in perforated domains, which are not covered by the homogenisation approximations. The meso-scale approximations presented here are uniform. Explicit asymptotic formulae are supplied with the remainder estimates.

\end{abstract}

\section{Introduction}

Meso-scale approximations have been 
introduced and rigorously studied in \cite{Maz_MN, Maz_MMS, MMN_Mesoelast}.
Physical applications in composite systems in electromagnetism were also addressed in the earlier papers \cite{Fig_1,Fig_2}.
The study of Green's kernels  as well as asymptotic analysis of solutions to eigenvalue problems for dense arrays of spherical obstacles was performed in \cite{Ozawa}.
Compared to classical homogenisation approaches (see \cite{Bakhvalov, SP, MarKhrus}), the  meso-scale approximation does not require any constraints on periodicity of the microstructure, and it is uniformly valid across  the whole domain, including neighbourhoods of singularly perturbed boundaries.  

Prior to the development of the meso-scale asymptotic approach, many papers and monographs 
(see, for example, \cite{Ciarlet, Ciarlet_Destuyunder, KMMI, KMMII}) have appeared which model singular perturbations of various domains. 
Examples include domains with irregular boundaries, thin components or domains containing either a single small defect or several defects.  The method of compound asymptotic expansions of solutions to such problems is described in \cite{OPTH1, OPTH2}. In particular, for domains with small defects, asymptotic 
approximations have proven to be 
superior to the finite element method (FEM), even when the 
overall number of defects is chosen to be 
large \cite{RAN}. 
For domains with perforations, the approximations presented in \cite{OPTH1, OPTH2} use model problems posed in the domain without defects  and problems posed in unbounded domains, in the exterior of individual inclusions. Integral characteristics of the defects are used here in connection with the energy of model fields in the exterior domains. 
For rigid inclusions we refer to the capacity of the inclusions, whereas  for voids we use the dipole matrix, that correspond to the Dirichlet and Neumann boundary conditions, respectively. 

The method of compound asymptotic expansions has also led to the development of uniform approximations for Green's kernels for domains with small defects for the Laplacian, corresponding to a variety of boundary value problems involving rigid inclusions \cite{CRM, JCOM}, voids \cite{Sob_vol} and soft inclusions \cite{AMS_tran}. Approximations for Green's kernels in long rods have also appeared in \cite{MMAS}. There also exist several approximations for Green's tensors of vector elasticity for solids with rigid inclusions \cite{RAN, AA} and holes with traction free boundaries \cite{MMN_book}.  Meso-scale approximations of Green's   function for the Laplacian in a solid with rigid boundaries has been derived in \cite{Maz_MN}. 

A systematic presentation of the theory of meso-scale approximations in densely perforated domains is given in the recent monograph \cite{MMN_book}. In particular, it was demonstrated that uniform meso-scale asymptotic approximations are of high importance for the analysis of fields in solids containing non-uniformly distributed clouds of small voids or inclusions. In such configurations, the traditional computational approaches like FEM are inefficient.  

\begin{figure}
\centering
\includegraphics[width=0.8\textwidth]{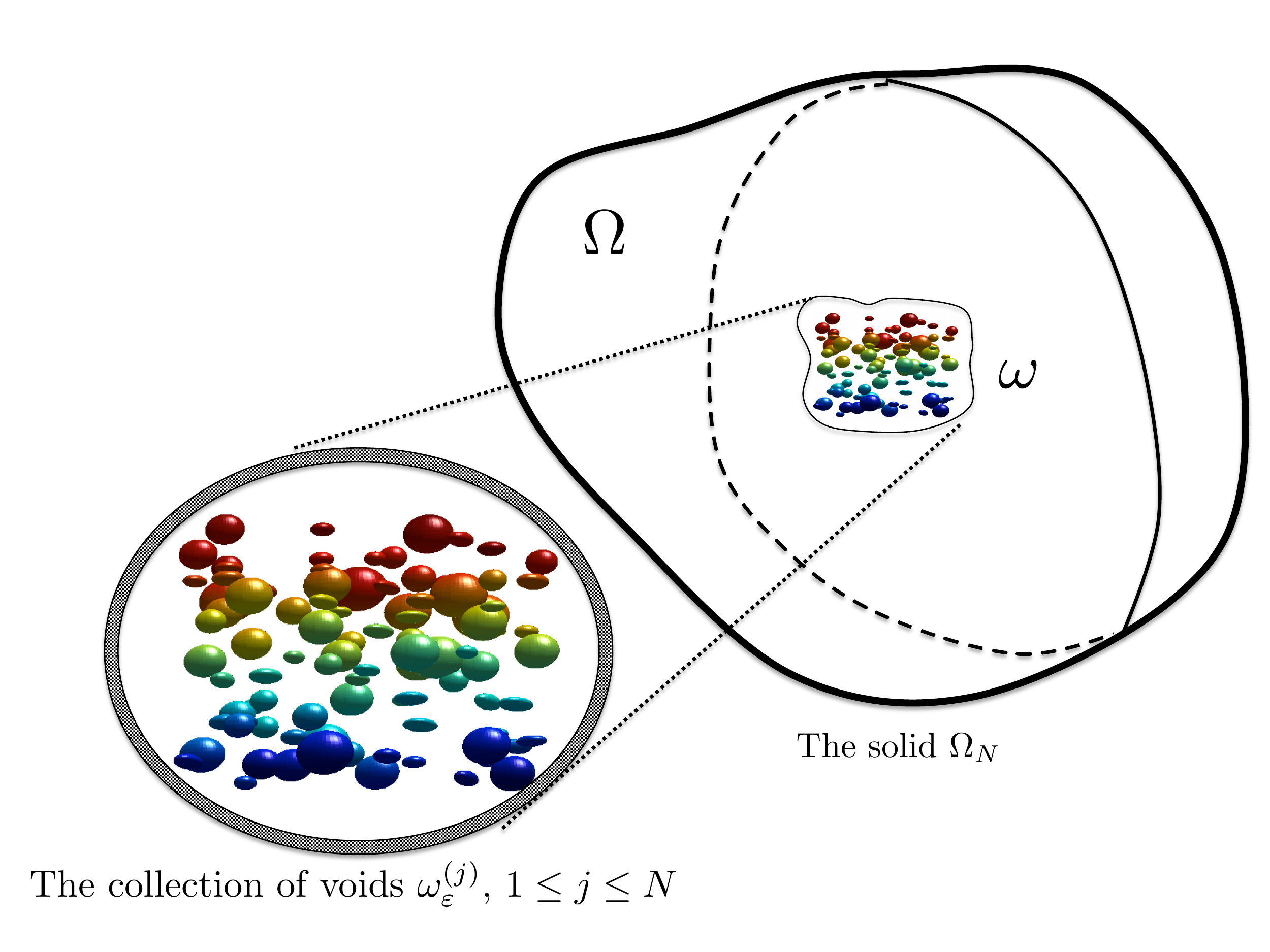}

\caption{The solid $\Omega_N$ containing a cloud $\omega$ of voids.}\label{OmN}
\end{figure}

Recently, the method used to develop meso-scale approximations   for scalar problems posed in solids with many small voids and inclusions has been extended to the Dirichlet problem of elasticity in solids with a cloud of rigid inclusions \cite{MMN_Mesoelast}. In the present paper the approach of \cite{MMN_book} is applied to a mixed boundary value problem of vector elasticity in an elastic solid, which contains a cloud of many voids whose boundaries are traction free. The number of voids is denoted by $N \gg 1.$ Each void is a concentrator of stress, and analysis of boundary layers is carried out in terms of special classes of dipole fields, which characterise the shape of voids and elastic properties of the material.  The schematic representation of the porous solids with a cloud of $N$ voids is shown in Fig. \ref{OmN}. Two small parameters are introduced as the normalised diameter of a  void and the minimal distance between neighbouring voids 
within the cloud.

Let $\Omega$ be a bounded domain in $\mathbb{R}^3$ representing an elastic solid. Contained in $\Omega$ are many small voids, $\omega^{(j)}_\varepsilon$, $1\le j\le N$ whose diameters are characterised by the small parameter $\varepsilon$ and that occupy a set $\omega \subset \Omega$ representing a cloud of voids. The sets $\Omega$ and $\omega^{(j)}_\varepsilon$, $j=1, \dots, N,$ are assumed to have  smooth boundaries. In addition, the minimum distance between the centres $\BO^{(k)}$, $1\le k \le N$, of  each void is connected with another small parameter $d$.
The geometry of the elastic solid with many small perforations will  be described  by the set $\Omega_N= \Omega \backslash \cup_{j=1}^N\overline{ \omega^{(j)}_\varepsilon}$. 

In the framework of vector elasticity, the Lam\'e operator  and the operator connected with the application of  external tractions will be denoted by $L(\nabla_\Bx)$ and  $T(\nabla_\Bx)$, respectively.

The displacement field $\Bu_N$ satisfies the governing equations of static elasticity:
\begin{equation}\label{equn1}
L(\nabla_{\Bx}) \Bu_N(\Bx)+\Bf(\Bx)=\BO\;, \quad \Bx \in \Omega_N\;,
\end{equation}
\begin{equation}\label{equn2}
\Bu_N(\Bx)=\BO\;, \quad \Bx\in \partial \Omega\;,
\end{equation}
\begin{equation}\label{equn3}
T_n(\nabla_\Bx)\Bu_N(\Bx)=\BO\;, \quad \Bx \in \partial \omega^{(j)}_\varepsilon\;, 1\le j \le N\;.
\end{equation}
In (\ref{equn1}), $\Bf \in L_\infty(\Omega_N)$ is a vector function representing the action of body forces inside the perturbed solid, that satisfies the constraints $\omega \cap \text{supp } \Bf  =\varnothing$ and $\text{dist}(\text{supp }\Bf, \partial \omega)\ge C$, with $C$ being a positive constant independent of $\varepsilon$ and $d$.

The construction of the  approximation for $\Bu_N$ presented here depends on several model fields:

\begin{enumerate}
\item the solution $\Bu$ of the problem in $\Omega$ without any voids,
\item the regular part $H$ of Green's tensor in $\Omega$,
\item a matrix function $\BQ^{(k)}$ that solves a Neumann problem in the exterior of the scaled void $\omega^{(k)}$ whose columns are known as the dipole fields for the elastic void; a re-scaling is applied to obtain $\BQ_\Gve^{(k)}$ for the small void $\omega_\Gve^{(k)}$
\item a constant matrix $\BM^{(k)}$, called the dipole matrix of the scaled void $\omega^{(k)}$, that characterises the void's shape and the elastic material properties.  The dipole matrix  $\BM^{(k)}_\varepsilon$ for the small void  $\omega_\varepsilon$ is constructed from $\BM^{(k)}$ by re-scaling.  The geometry of the voids  is assumed to be chosen so that the maximum and minimum eigenvalues $\lambda^{(k)}_{\text{max}}$ and $\lambda^{(k)}_{\text{min}}$, respectively, of the matrix $-\BM^{(k)}_\varepsilon$ satisfy the inequalities
\begin{equation}\label{eigvalM}
C_1\, \varepsilon^3 \le \lambda^{(k)}_{\text{min}}\qquad  \text{ and } \qquad \lambda^{(k)}_{\text{max}}\le C_2\, \varepsilon^3\;,
\end{equation}
for $k=1,\dots, N$, where $C_1$ and $C_2$ represent different positive constants.
\end{enumerate}

For convenience of notations, we also use the vector $\BE$ of normalised elastic strain, corresponding to the displacement field $\Bu$,  so that 
$\BE(\Bu)=\BGX(\nabla_\Bx)\Bu,$ where $\BGX$ is the linear matrix differential operator. 


The constant vector $\BV$ and matrices $\BM$ and $\BS$ are also 
used in the approximation for $\Bu_N$:
\[\BV=((\BGX(\nabla_\Bx)^T\Bu(\Bx))^T\Big|_{\Bx=\BO^{(1)}},\dots, (\BGX(\nabla_\Bx)^T\Bu(\Bx))^T\Big|_{\Bx=\BO^{(N)}})^T\;, \]
\[\BM=\text{diag}\{\BM_\varepsilon^{(1)},\dots, \BM_\varepsilon^{(N)}\}\]
and
\[\BS=\left\{\begin{array}{ll}\BGX(\nabla_{\Bx})^T(\BGX(\nabla_\By)^TG(\By, \Bx))^T\Big|_{\substack{\Bx=\BO^{(i)}\\ \By=\BO^{(j)}}}
\;,&\quad \text{ if }\quad i\ne j\;,\\
\mathbb{O}_{6\times 6}\;,&\quad \text{otherwise}\;,
\end{array}\right.\]
where $\mathbb{O}_{6\times 6}$ is the $6\times 6$ null  matrix; also in the text below  $\mathbb{I}_{n\times n}$ will stand for the $n\times n$ identity  matrix.

The {\bf main result} of this article is the uniform asymptotic approximation of the displacement field $\Bu_N$, as presented in the following theorem: 

\begin{thm}\label{th1}
Let the small parameters $\varepsilon$ and $d$ satisfy the inequality
\begin{equation}\label{constraint1}
\varepsilon < c\, d\;,
\end{equation}
with $c$ being a sufficiently small constant. Then the approximation for $\Bu_N$ is given by 
\begin{equation}\label{uNapprox}
\Bu_N(\Bx)=\Bu(\Bx)+\sum_{k=1}^N\Big\{\BQ^{(k)}_\varepsilon(\Bx)-(\BGX(\nabla_\Bz)^TH(\Bz, \Bx))^T\BM_\varepsilon^{(k)}\Big|_{\Bz=\BO^{(k)}}\Big\}\BC^{(k)}+\BR_N(\Bx)
\end{equation}
where $\BC=((\BC^{(1)})^T, \dots, (\BC^{(N)})^T)^T$ solves the linear algebraic system
\begin{equation}\label{alg_sys_eq}
-\BV=(\mathbb{I}_{6N\times 6N}+\BS\BM)\BC\;,
\end{equation}
and for the remainder $\BR_N$, the energy estimate holds
\begin{equation}\label{eqestRN}
\int_{\Omega_N}\text{\emph{tr}}(\BGs(\BR_N)\Be(\BR_N))d\Bx\le  \text{\emph{Const }}\Big\{\varepsilon^{11}d^{-11}+\varepsilon^5d^{-3}\Big\}\|\BE(\Bu)\|^2_{L_\infty(\Omega)}\;.
\end{equation}
Here $\text{\emph{Const}}$ in the above right-hand side is independent of $\varepsilon$ and $d$.
\end{thm}

This representation (\ref{uNapprox}) is uniform and it engages several classes of model fields, which are independent of the small parameters $\varepsilon$ and $d$ (also see \cite{MMN_book}).

The structure of the paper is as follows. 
Main notations are introduced in Section \ref{geometry}. 
Model problems used to approximate $\Bu_N$ are introduced in Section \ref{mod_field}. The formal approximation of $\Bu_N$ is then provided in Section \ref{formal_app}. 
This approximation relies on the solution of the algebraic system (\ref{alg_sys_eq}) and the solvability of this system is studied under the constraint (\ref{constraint1}) in Section \ref{alg_sys_inv}. Then, in Section \ref{en_est_RN}, the energy estimate (\ref{eqestRN}) for the remainder of the approximation is proved. Simplified asymptotic approximations for $\Bu_N$ are then given in Section \ref{simp_asy}. Following this, conclusions and discussion are given in Section \ref{conclusions}. Appendix A contains a local regularity estimate used in the proof of the energy estimate  (\ref{eqestRN}). In Appendix B, a detailed proof of intermediate steps used to show the solvability of  (\ref{alg_sys_eq}) is presented. Finally, in Appendix C, we show that for certain geometries, dipole characteristics can be constructed in the closed form for the case of spherical cavities and explicit representations are given. 

\section{Geometry of the perforated domain and main notations} \label{geometry}
A domain $\Omega \subset \mathbb{R}^3$ will be used to denote the set corresponding to an elastic solid without holes, with smooth frontier $\partial \Omega$. For a small positive parameter $\varepsilon>0$,  the open set $\omega^{(j)}_\varepsilon$  is defined in such a way that it contains an interior  point $\BO^{(j)}$,  has  smooth boundary $\partial \omega^{(j)}_\varepsilon$ and a diameter characterised by $\varepsilon$. The collection of sets $\omega^{(j)}_\varepsilon$, $1\le j \le N$, will represent the small voids contained inside the set $\Omega$ that are subject to some further geometric constraints discussed below. In this way, we define the perturbed geometry $\Omega_N=\Omega\backslash \cup_{j=1}^N \overline{\omega^{(j)}_\varepsilon}$. 
It is also assumed that a small parameter $d$ characterises the minimum distance between points in the array $\{\BO^{(j)}\}_{j=1}^N$, and that this minimum distance is $2d$. Another geometric constraint is the assumption of the existence of a set $\omega$ that satisfies 
\[\bigcup_{j=1}^N\omega^{(j)}_\varepsilon \subset \omega, \quad \text{dist}(\bigcup_{j=1}^N\omega^{(j)}_\varepsilon, \partial \omega )\ge 2d\quad \text{ and }\quad \text{dist}(\partial \omega,\partial \Omega)\ge 1\;. \]
It is also useful to introduce the matrix functions:
\begin{equation}\label{int1}
\BGX(\Bx)=\left(\begin{array}{ccccccccccc}
x_1 & & 0 && 0&& 2^{-1/2} x_2&& 2^{-1/2} x_3&&0\\
0&&x_2 && 0&&2^{-1/2} x_1&& 0&& 2^{-1/2} x_3\\
0&& 0 &&x_3 &&0&&2^{-1/2} x_1&&2^{-1/2} x_2
\end{array}\right)
\end{equation}
and
\begin{equation}\label{int2}
\BGx(\Bx)=\left(\begin{array}{ccccccccccc}
1 & & 0 && 0&& 2^{-1/2} x_2&& 2^{-1/2} x_3&&0\\
0&&1 && 0&&-2^{-1/2} x_1&& 0&& 2^{-1/2} x_3\\
0&& 0 &&1 &&0&&-2^{-1/2} x_1&&-2^{-1/2} x_2
\end{array}\right)\;.
\end{equation}
These matrices satisfy the conditions
\[\BGX(\nabla_\Bx)^T\BGX(\Bx)=\mathbb{I}_{6\times 6} \qquad  \BGX(\nabla_\Bx)^T\BGx(\Bx)=\mathbb{O}_{6\times 6}\;,\]
where $\mathbb{I}_{n \times n}$ and $\mathbb{O}_{n \times n}$ are the $n\times n$ identity and null matrices, respectively.

The matrices $\BGx$ and $\BGX$  also lead to a compact form of the first-order Taylor approximation for a vector function $\Bu$ about $\Bx=\BO$
\[\Bu(\BO)=\BGx(\Bx)\BGx(\nabla_\Bx)^T\Bu(\BO)+\BGX(\Bx)\BGX(\nabla_\Bx)^T\Bu(\BO)
+O(|\Bx|^2)\;,
\]
and allow the Lam\'e operator $L(\nabla_\Bx)$ to be defined as
\[L(\nabla_\Bx):=\BGX(\nabla_{\Bx})\BA\BGX(\nabla_{\Bx})^T\;,\]
with 
\[\BA=\left(\begin{array}{ccc}
\BB&&\mathbb{O}_{3\times 3}\\ \\
\mathbb{O}_{3\times 3}&& 2\mu \mathbb{I}_3\end{array}\right)\;,
 \qquad \BB=\left(\begin{array}{ccccc}\lambda+2\mu && \lambda && \lambda\\
\lambda&& \lambda+2\mu  && \lambda\\
\lambda&& \lambda  && \lambda+2\mu
\end{array}\right)\;.\]
The corresponding traction operator $T_n(\nabla_\Bx)$ is then
\[T_n(\nabla_\Bx):=\BGX(\Bn)\BA\BGX(\nabla_{\Bx})^T\;,\]
which will be applied on the boundary of an open set  with  $\Bn$ being the unit-outward normal to the  set.

The  strain tensor $\Be(\Bv)=[e_{ij}(\Bv)]_{i,j=1}^3$, stress tensor $\BGs(\Bv)=[\sigma_{ij}(\Bv)]_{i,j=1}^3$ and the tensor of rotations $\BGn(\Bv)=[\eta_{ij}(\Bv)]_{i,j=1}^3$ for a vector field $\Bv$ take the forms
\[\Be(\Bv)=\frac{1}{2}((\nabla \otimes \Bv)+(\nabla \otimes \Bv)^T)\;, \quad \BGs(\Bv)=\lambda\text{tr}(\Be(\Bv)) \mathbb{I}_3+2\mu\Be(\Bv)\;,\]
and
\[\BGn(\Bv)=\frac{1}{2}((\nabla \otimes \Bv)-(\nabla \otimes \Bv)^T)\;.\]
The matrix $\BJ=[J^{(i)}]_{i=1}^3$, where $J^{(i)}$ is the $i^{th}$ column of this matrix, is
\begin{equation}
\BJ(\Bx)=\left(\begin{array}{ccccc}
0     & & -x_3  & &x_2\\
x_3 & &  0     & &-x_1\\
-x_2 & & x_1 & & 0 
\end{array}\right)\;,
\end{equation}
and this plays a role in the description of the overall moment acting on an elastic  body. It is noted that \[\BGn(J^{(1)})=\left(\begin{array}{ccccc}
0 && 0&& 0\\
0&&0&&1\\
0&&-1&&0
\end{array}\right)\;, \quad\BGn(J^{(2)})=\left(\begin{array}{ccccc}
0 && 0&&-1\\
0&& 0&& 0\\
1&& 0&& 0
\end{array}\right)\quad \text{ and }\quad \BGn(J^{(3)})=\left(\begin{array}{ccccc}
0 && 1&&0\\
-1&& 0&&0\\
0&& 0&&0
\end{array}\right) \;.\]
The strain and stress vectors denoted by $\BE$  and $\BN$, respectively, are  defined by 
\begin{equation}\label{EandN}\begin{array}{c}
\BE=(e_{11}, e_{22}, e_{33}, \sqrt{2}e_{12}, \sqrt{2}e_{13}, \sqrt{2}e_{23})^T\;,\\ \\\BN=(\sigma_{11}, \sigma_{22}, \sigma_{33}, \sqrt{2}\sigma_{12}, \sqrt{2}\sigma_{13}, \sqrt{2}\sigma_{23})^T\;,
\end{array}
\end{equation}
and can also be introduced through the matrix operator (\ref{int1})  as 
\begin{equation}\label{eqXISIG}
\BE(\Bv)=\BGX(\nabla_\Bx)\Bv\qquad \BN(\Bv)=\BA\BGX(\nabla_\Bx)\Bv\;, 
\end{equation}
for a vector function $\Bv$.
Note also that the quantity $S(\BU)=\text{tr}(\Be(\BU)\Be(\BU))$ can also be represented as
\begin{equation}\label{eqtreE}
S(\BU)
=\BE(\BU)^T\left(\begin{array}{ccc}
\mathbb{I}_3&&\mathbb{O}_3\\ \\
\mathbb{O}_3&& 2^{-1} \mathbb{I}_3\end{array}\right)\BE(\BU)\;.
\end{equation}

\section{Model  fields}\label{mod_field}
In this section, we discuss the model fields used in the meso-scale approximation of $\Bu_N$ in detail. We begin with an introduction of fields defined in the unperturbed set $\Omega$:
\begin{enumerate}
\item \emph{The solution of the exterior Dirichlet problem.} The vector field $\Bu$ is a solution of 
\begin{equation}\label{up1}
L(\nabla_\Bx)\Bu(\Bx)+\Bf(\Bx)=\BO\;, \quad \Bx\in \Omega\;,
\end{equation} 
\begin{equation}\label{upbc}
\Bu(\Bx)=\BO\;,\quad \Bx \in \partial \Omega\;,
\end{equation}
where $\Bf$ satisfies the same conditions as in the statement of problem (\ref{equn1})--(\ref{equn3}), and the same notation will be used to represent the extension of $\Bf$ by  zero inside the voids $\omega_\varepsilon^{(j)}$, $1\le j \le N$.
\item \emph{The Green's tensor for the solid $\Omega$.} The notation $G$ will refer to the Green's tensor in the domain $\Omega$, that is a solution of 
\begin{equation}\label{Gp1}
L(\nabla_{\Bx})G(\Bx, \By)+\delta(\Bx-\By)\mathbb{I}_3=0\mathbb{I}_3\;, \quad \Bx, \By \in \Omega\;,
\end{equation}
and satisfies the homogeneous Dirichlet condition 
\begin{equation}\label{Gp2}
G(\Bx,\By)=0\mathbb{I}_3\;, \quad \Bx \in \partial \Omega, \By \in \Omega\;.
\end{equation}
The regular part $H$ of this tensor is represented by
\begin{equation}
\label{Gp3}
H(\Bx, \By)=\Gamma(\Bx, \By)-G(\Bx, \By)\;,
\end{equation}
where $\Gamma$ is the Kelvin-Somigliana tensor
\begin{equation}\label{Gamma_matrix}
\Gamma(\Bx,\By)=\frac{1}{8\pi\mu(\lambda+2\mu)}\frac{1}{|\Bx-\By|}\left\{(\lambda+3\mu)\mathbb{I}_3+(\lambda+\mu)\frac{(\Bx-\By)\otimes(\Bx-\By)}{|\Bx-\By|^2}\right\}\;,
\end{equation}
and 
\[L(\nabla_\Bx)\Gamma(\Bx,\By)+\delta(\Bx-\By)\mathbb{I}_3=0\mathbb{I}_3\;.\] 
The above problem then implies that $H(\Bx, \By)=(H(\By, \Bx))^T$, $\Bx, \By\in \Omega$.

Next, we introduce the boundary layer fields for the small voids, known as the dipole fields \cite{Sob_vol, MMP}:
\item \emph{The dipole fields for the voids. } In the construction of the boundary layers for the asymptotic algorithm in the vicinity of the void $\omega^{(j)}_\varepsilon$, the physical fields known as dipole fields, will play an essential role. They are defined as functions of the scaled variable $\BGx_j=\varepsilon^{-1}(\Bx-\BO^{(j)})$ outside of the scaled set $\omega^{(j)}=\{\BGx_j: \varepsilon\BGx_j+\BO^{(j)}\in \omega^{(j)}_\varepsilon\}$. The dipole fields form the columns of the $3\times 6$ matrix  $\BQ^{(j)}$ where
\begin{equation}\label{Qp1}
L(\nabla_{\BGx_j})\BQ^{(j)}({\BGx_j})=\mathbb{O}_{3\times 6}\;, \quad \BGx_j\in\mathbb{R}^3\backslash  \overline{\omega^{(j)}}\;,
\end{equation}
\begin{equation}\label{Qp2}
T_n(\nabla_{\BGx_j})\BQ^{(j)}({\BGx_j})=\BGX(\Bn^{(j)})\BA\;,\quad \BGx_j\in \partial \omega^{(j)}\;, 
\end{equation}
\begin{equation}\label{Qp3}
\BQ^{(j)} ({\BGx_j})\to\mathbb{O}_{3\times 6}\;,\text{ as } \quad |{\BGx_j}| \to \infty\;,
\end{equation}
where $\Bn^{(j)}$ is the unit outward normal to $\mathbb{R}^3\backslash \overline{\omega^{(j)}}$.
\end{enumerate}

The right-hand sides in the Neumann boundary condition (\ref{Qp2}) are subjected to the constraints that the total force on boundary $\partial \omega^{(j)}$ and the resultant moments are zero:
\begin{equation}\label{Qp4}
\int_{\partial\omega^{(j)}}T_n(\nabla_{\scriptsize \BGx_j})\BQ^{(j)}({\BGx_j})ds_{\scriptsize \BGx_j}=\mathbb{O}_{3\times 6}\;,
\end{equation}
\begin{equation}\label{Qp5}
\int_{\partial\omega^{(j)}}\BJ(\BGx_j) T_n(\nabla_{\scriptsize \BGx_j})\BQ^{(j)}({\BGx_j})ds_{\scriptsize \BGx_j}=\mathbb{O}_{3\times 6}\;.
\end{equation}
A special matrix  $\BM^{(j)}$, with constant entries,  is also required to construct the leading order behaviour of the matrix $\BQ^{(j)}$ at infinity and this is called the dipole matrix. The behaviour of $\BQ^{(j)}$ far away from the void $\omega^{(j)}$ is described in the next lemma (see {\cite{Sob_vol, MMP}}). 

\begin{lem}\label{lemQ}
 For $|\BGx_j|>2$ the matrix $\BQ^{(j)}$ admits the form
\[\BQ^{(j)}(\BGx_j)=-(\BGX(\nabla_{\BGx_j})^T\Gamma(\BGx_j, \BO ))^T\BM^{(j)}+O(|\BGx_j|^{-3})\;.\]
\end{lem}

\section{Formal meso-scale approximation for $\Bu_N$}\label{formal_app}
In this section, the derivation of the meso-scale asymptotic approximation for $\Bu_N$ in Theorem \ref{th1} is formally derived. 
First we note that in what follows, we will need the matrices $\BQ_\varepsilon^{(j)}(\Bx)=\varepsilon \BQ^{(j)} (\BGx_j)$ and $\BM^{(j)}_\varepsilon=\varepsilon^3 \BM^{(j)}$.
 According to \cite{Sob_vol, MMP}, the dipole matrix $\BM^{(j)}$
  is symmetric negative definite. 
  
  \vspace{0.1in}In the next Lemma and the following text the notation $\text{Const}$ will represent different positive constants independent of the  parameters $\varepsilon$, $d$ and $N$.


\vspace{0.1in}The meso-scale approximation for the displacement field $\Bu_N$ is now defined by
\begin{lem}\label{lemuNapp}
The formal approximation of  $\Bu_N$  is given in the form
\begin{equation}\label{thm1}
\Bu_N(\Bx)=\Bu(\Bx)+\sum_{k=1}^N\Big\{\BQ^{(k)}_\varepsilon(\Bx)-(\BGX(\nabla_\Bz)^TH(\Bz, \Bx))^T\BM_\varepsilon^{(k)}\Big|_{\Bz=\BO^{(k)}}\Big\}\BC^{(k)}+\BR_N(\Bx)
\end{equation}
where the coefficients $\BC^{(j)}$ satisfy 
\begin{equation}\label{algeq}
\BGX(\nabla_{\Bx})^T\Bu(\Bx)\Big|_{\Bx=\BO^{(j)}}+\BC^{(j)}+\sum_{\substack{k\ne j\\ 1\le k\le N}}\BGX(\nabla_{\Bx})^T(\BGX(\nabla_\Bz)^TG(\Bz, \Bx))^T\BM_\varepsilon^{(k)}\Big|_{\substack{\Bz=\BO^{(k)}\\
\Bx=\BO^{(j)}}}\BC^{(k)}=\BO\;,
\end{equation}
for $1\le j \le N$.
The remainder $\BR_N$ is a solution of the boundary value problem 
for the homogeneous Lam\'e equation 
in $\Omega_N$,  
with the mixed boundary conditions:
$$
\BR_N(\Bx) = \BGf(\Bx) ~ \mbox{on }~ \prt \GO; ~~~\mbox{\rm and} ~~
T_n(\nabla_{\Bx})\BR_N(\Bx) = \BGf^{(j)} (\Bx) ~ \mbox{on} ~ \Bx \in \partial \omega^{(j)}_\varepsilon\;, 1\le j\le N,
$$
where the right-hand sides satisfy the estimates
\begin{equation}
|  \BGf(\Bx)| \le \text{\emph{Const} }\sum_{k=1}^N\frac{\varepsilon^4|\BC^{(k)}|}{|\Bx-\BO^{(k)}|^3}\;, \qquad \Bx\in\partial \Omega\;, 
\end{equation} 
and
\begin{eqnarray}
|\BGf^{(j)}(\Bx)   |\le \text{\emph{Const} }
\Bigg(\varepsilon(1+\varepsilon^2|\BC^{(j)}|)+\sum_{\substack{k\ne j\\ 1\le k\le N}}\frac{\varepsilon^4|\BC^{(k)}|}{|\Bx-\BO^{(k)}|^4}\Bigg)\;, \quad \Bx \in \partial \omega^{(j)}_\varepsilon\;, 1\le j\le N\;,
\end{eqnarray}
and $\BGf^{(j)}$, $1\le j \le N$, fulfil the orthogonality conditions
\begin{equation}
\label{othocond}
\int_{\partial\omega^{(j)}_\varepsilon}\BGf^{(j)}ds_\Bx=\BO\;,\quad 
\int_{\partial\omega^{(j)}_\varepsilon}\BJ(\Bx-\BO^{(j)}) \BGf^{(j)}ds_\Bx=\BO\;, \quad 1\le j \le N\;.
\end{equation}
\end{lem}

\emph{Proof.} The orthogonality conditions (\ref{othocond}) follow from (\ref{thm1}), the Betti formula 
 and the model problems introduced in Section 2. 

According to problem 1, Section 2,  the vector function 
\begin{equation}\label{alg1a}
{\BR}_N^{(1)}=\Bu_N(\Bx)-\Bu(\Bx)\;,
\end{equation}
satisfies the homogeneous Lam\'e equation for $\Bx\in \Omega_N$. Since both $\Bu_N(\Bx)$ and $\Bu(\Bx)$ satisfy the homogeneous Dirichlet condition on $\partial \Omega$, then $\BR^{(1)}_N(\Bx)=\BO$, for $\Bx\in \partial \Omega$. Next consider the tractions of the $\BR^{(1)}_N$ on $\partial \omega^{(j)}_\varepsilon$. This condition, using Taylor's expansion about $\Bx=\BO^{(j)}$ takes the form
\begin{eqnarray}
T_n(\nabla_{\Bx})\BR_N^{(1)}&=&T_n(\nabla_{\Bx})(\Bu_N(\Bx)-\Bu(\Bx))=-T_n(\nabla_{\Bx})\Bu(\Bx)\;,\nonumber\\
&=&-T_n(\nabla_{\Bx})\Bu(\Bx)\Big|_{\Bx=\BO^{(j)}}+O(\varepsilon)\;, \quad \Bx \in \partial \omega^{(j)}_\varepsilon\;, 1\le j\le N\;.\label{alg1}
\end{eqnarray}
An approximation for $\BR^{(1)}_N$ is then sought as
\begin{equation}\label{R1app}
\BR^{(1)}_N(\Bx)=\sum_{k=1}^N\Big\{\BQ^{(k)}_\varepsilon(\Bx)-(\BGX(\nabla_\Bz)^TH(\Bz, \Bx))^T\BM_\varepsilon^{(k)}\Big|_{\Bz=\BO^{(k)}}\Big\}\BC^{(k)}+\BR_N(\Bx)\;.
\end{equation}
The goal is now to determine the vector coefficients $\BC^{(k)}$, $1\le k\le N$, to complete the formal approximation. It is noted that the remainder in (\ref{R1app}) is a solution of 
\[L(\nabla_{\Bx})\BR_N(\Bx)=\BO\;, \qquad \Bx\in \Omega_N\;,\]
and from the boundary condition for regular part $H$ of Green's tensor (see  (\ref{Gp1})--(\ref{Gp3})), the exterior Dirichlet condition for $\BR_N$ is
\begin{eqnarray}
\BR_N(\Bx)&=&-\sum_{k=1}^N\Big\{\BQ^{(k)}_\varepsilon(\Bx)-(\BGX(\nabla_\Bz)^T\Gamma(\Bz, \Bx))^T\BM_\varepsilon^{(k)}\Big|_{\Bz=\BO^{(k)}}\Big\}\BC^{(k)}\nonumber\\
&=&O\left(\sum_{k=1}^N\frac{\varepsilon^4|\BC^{(k)}|}{|\Bx-\BO^{(k)}|^3}\right)\;, \quad \Bx \in \partial \Omega\;, \label{exbc1}
\end{eqnarray}
where Lemma \ref{lemQ} has also been used.
In order to derive the vector coefficients $\BC^{(j)}$, $1\le j \le N$, the tractions on the interior boundaries for $\BR_N$ should be considered. For $\Bx \in \partial \omega^{(j)}_\varepsilon$, according to (\ref{alg1})
\begin{eqnarray}
T_n(\nabla_{\Bx})\BR_N(\Bx)&=&-T_n(\nabla_{\Bx})\Bu(\Bx)\Big|_{\Bx=\BO^{(j)}}-T_n(\nabla_{\Bx})\BQ^{(j)}_\varepsilon(\Bx)\BC^{(j)}\nonumber \\
&&- \sum_{\substack{k\ne j\\ 1\le k\le N}}T_n(\nabla_{\Bx})\Big\{\BQ^{(k)}_\varepsilon(\Bx)-(\BGX(\nabla_\Bz)^TH(\Bz, \Bx))^T\BM_\varepsilon^{(k)}\Big|_{\Bz=\BO^{(k)}}\Big\}\BC^{(k)}\nonumber \\
&&
+O(\varepsilon)+O(\varepsilon^3|\BC^{(j)}|)\;, \quad \Bx \in \partial \omega^{(j)}_\varepsilon\;, 1\le j\le N\;.\nonumber
\end{eqnarray}
Condition (\ref{Qp2}) and Lemma \ref{lemQ} then provide a simplified form of the above traction condition on $\partial \omega^{(j)}_\varepsilon$:
\begin{eqnarray}
T_n(\nabla_{\Bx})\BR_N(\Bx)&=&-\BGX(\Bn^{(j)})\BA\Big\{\BGX(\nabla_{\Bx})^T\Bu(\Bx)\Big|_{\Bx=\BO^{(j)}}+\BC^{(j)}\nonumber \\
&&+ \sum_{\substack{k\ne j\\ 1\le k\le N}}\BGX(\nabla_{\Bx})^T(\BGX(\nabla_\Bz)^TG(\Bz, \Bx))^T\BM_\varepsilon^{(k)}\Big|_{\Bz=\BO^{(k)}}\BC^{(k)}\Big\}\nonumber \\
&&
+O(\varepsilon)+O(\varepsilon^3|\BC^{(j)}|)+O\Bigg(\sum_{\substack{k\ne j\\ 1\le k\le N}}\frac{\varepsilon^4|\BC^{(k)}|}{|\Bx-\BO^{(k)}|^4}\Bigg)\;, \quad \Bx \in \partial \omega^{(j)}_\varepsilon\;, 1\le j\le N\;.\nonumber
\end{eqnarray}
Applying the Taylor expansion once more about $\Bx=\BO^{(j)}$ gives
\begin{eqnarray}
&&T_n(\nabla_{\Bx})\BR_N(\Bx)\nonumber\\
&=&-\BGX(\Bn^{(j)})\BA\Big\{\BGX(\nabla_{\Bx})^T\Bu(\Bx)\Big|_{\Bx=\BO^{(j)}}+\BC^{(j)}+\sum_{\substack{k\ne j\\ 1\le k\le N}}\BGX(\nabla_{\Bx})^T(\BGX(\nabla_\Bz)^TG(\Bz, \Bx))^T\BM_\varepsilon^{(k)}\Big|_{\substack{\Bz=\BO^{(k)}\\
\Bx=\BO^{(j)}}}\BC^{(k)}\Big\}\nonumber \\
&&
+O(\varepsilon)+O(\varepsilon^3|\BC^{(j)}|)+O\Bigg(\sum_{\substack{k\ne j\\ 1\le k\le N}}\frac{\varepsilon^4|\BC^{(k)}|}{|\BO^{(j)}-\BO^{(k)}|^4}\Bigg)\;, \quad \Bx \in \partial \omega^{(j)}_\varepsilon\;, 1\le j\le N\;.
\end{eqnarray}
Thus, we can remove the leading order discrepancy in the preceding boundary condition by allowing $\BC^{(j)}$ to satisfy the system of equations 
\begin{eqnarray}
&&\BGX(\nabla_{\Bx})^T\Bu(\Bx)\Big|_{\Bx=\BO^{(j)}}+\BC^{(j)}+\sum_{\substack{k\ne j\\ 1\le k\le N}}\BGX(\nabla_{\Bx})^T(\BGX(\nabla_\Bz)^TG(\Bz, \Bx))^T\BM_\varepsilon^{(k)}\Big|_{\substack{\Bz=\BO^{(k)}\\
\Bx=\BO^{(j)}}}\BC^{(k)}=\BO\;,\label{system1}
\end{eqnarray}
for $1\le j \le N$. Combining (\ref{alg1a}), (\ref{R1app}) and (\ref{exbc1})--(\ref{system1})  completes the proof of the lemma. \hfill $\Box$

\section{Algebraic system for $\BC^{(j)}$ and  its solvability}\label{alg_sys_inv}
Before presenting the energy estimate for the remainder $\BR_N$, the solvability of the algebraic system (\ref{algeq}) is discussed in this section under the constraint that $\varepsilon <c\, d$. We first introduce some notations to simplify the analysis.

Using the following vectors,
\[\BC=((\BC^{(j)})^T, \dots, (\BC^{(N)})^T)^T\quad \text{ and }\quad \BV=((\BGX(\nabla_\Bx)^T\Bu(\Bx))^T\Big|_{\Bx=\BO^{(1)}},\dots, (\BGX(\nabla_\Bx)^T\Bu(\Bx))^T\Big|_{\Bx=\BO^{(N)}})^T\;,  \]
and the $6N\times 6N$ symmetric matrices:
\[\BM=\text{diag}\{\BM_\varepsilon^{(1)},\dots, \BM_\varepsilon^{(N)}\}\;,\]
\[\BS=\left\{\begin{array}{ll}
\BGX(\nabla_{\Bx})^T(\BGX(\nabla_\By)^TG(\By, \Bx))^T\Big|_{\substack{\Bx=\BO^{(j)}\\ \By=\BO^{(k)}}}\;,&\quad \text{ if }\quad j\ne k\;,\\
\mathbb{O}_{6\times 6}\;,&\quad{\text{ otherwise}}\;.
\end{array}\right.\]
the  equations (\ref{algeq}) can be written as
\begin{equation}\label{inalgsys1}
-\BV=(\mathbb{I}_{6N\times 6N}+\BS\BM)\BC\;.
\end{equation}

\subsection{Solvability of the algebraic system (\ref{inalgsys1})}
Here, a result concerning  the solvability of the system (\ref{inalgsys1}) is proved:
\begin{lem}\label{lem_invert}
Let the parameters $\varepsilon$ and $d$ satisfy the inequality
\begin{equation}\label{eqcd}
\varepsilon <c d\;,
\end{equation}
where $c$ is a sufficiently small constant.
Then, the linear algebraic system $(\ref{inalgsys1})$ is solvable and 
\begin{equation}\label{invertest}
\sum_{j=1}^N |\BC^{(j)}|^2 \le \text{\emph{Const }}\sum_{j=1}^N |\BE(\Bu(\Bx))|^2\Big|_{\Bx=\BO^{(j)}}\;,
\end{equation}
where the strain vector $\BE(\Bu(\Bx))$ is defined in $(\ref{EandN})$.
\end{lem}

\vspace{0.1in}\emph{Proof. }By taking the  scalar product of (\ref{inalgsys1}) with $\BM \BC$ and using the Cauchy inequality we deduce
\begin{equation}\label{est_e_d_1}
\langle -\BM\BC,\BC\rangle-\langle \BM\BC, \BS\BM\BC \rangle=\langle \BM \BC, \BV\rangle \le \langle -\BM\BC,\BC\rangle^{1/2}\langle-\BM\BV,\BV\rangle^{1/2} \; .
\end{equation}

Note that the term $\langle \BM\BC, \BS\BM\BC \rangle$ admits the form
\begin{equation}
\label{SP_1}
\langle \BM\BC, \BS\BM\BC \rangle= \sum_{j=1}^N (\BM_\varepsilon^{(j)}\BC^{(j)})^T \cdot \sum_{\substack{k\ne j\\1\le k \le N}}
\BGX(\nabla_{\Bx})^T(\BGX(\nabla_\By)^TG(\By, \Bx))^T\Big|_{\substack{\Bx=\BO^{(j)}\\ \By=\BO^{(k)}}}
(\BM_\varepsilon^{(k)}\BC^{(k)}) \;.
\end{equation}

In Appendix B, 
it is shown that
 (\ref{SP_1})
 satisfies
\begin{equation}\label{QF}
|\langle \BM\BC, \BS\BM\BC \rangle| \le \text{Const }d^{-3}\langle\BM\BC, \BM\BC\rangle\;.
\end{equation}
Returning to (\ref{est_e_d_1}), this can then be used to establish that
\begin{eqnarray*}
\langle -\BM \BV,\BV \rangle^{1/2} &\ge& \langle -\BM \BC,\BC \rangle^{1/2} -\frac{\langle\BM\BC, \BS\BM\BC\rangle}{\langle -\BM \BC,\BC \rangle^{1/2} }\\
&\ge& \langle -\BM \BC,\BC \rangle^{1/2} -\text{Const }d^{-3}\frac{\langle\BM\BC, \BM\BC\rangle}{\langle -\BM \BC,\BC \rangle^{1/2} } \;.
\end{eqnarray*}
We note that 
\[\langle\BM\BC, \BM\BC\rangle=\langle-\BM\BC, -\BM\BC\rangle\le \text{Const }\max_{1\le k\le N}\lambda_{\text{max}}^{(k)}\; \langle-\BM\BC, \BC\rangle \]
and since the eigenvalues of the dipole matrices $-\BM^{(k)}_\varepsilon$, $1\le k \le N$ are $O(\varepsilon^3)$ according to (\ref{eigvalM}), it follows 
\begin{eqnarray*}
\langle -\BM \BV,\BV \rangle^{1/2} &\ge&(1-\text{Const }\varepsilon^3 d^{-3})\langle -\BM \BC,\BC \rangle^{1/2}\;.
\end{eqnarray*}
Estimate (\ref{invertest}) now follows from (\ref{eqcd}) and (\ref{eqXISIG}). The proof is complete. \hfill $\Box$
\section{Energy estimate for the remainder $\BR_N$}\label{en_est_RN}

With the formal meso-scale asymptotic approximation of $\Bu_N$ in place,  the energy estimate for the remainder  term $\BR_N$  in Theorem \ref{th1}  is now obtained.

\begin{lem}\label{lemenest}
Let the parameters $\varepsilon$ and $d$ satisfy the inequality
\[\varepsilon<c\,d\]
where $c$ is a sufficiently small constant. Then the remainder term $\BR_N$ satisfies the energy estimate
\begin{equation}\label{enesR}
\int_{\Omega_N}\text{\emph{tr}}(\BGs(\BR_N)\Be(\BR_N))d\Bx\le  \text{\emph{Const }}\Big\{\varepsilon^{11}d^{-11}+\varepsilon^5d^{-3}\Big\}\|\BE(\Bu)\|^2_{L_\infty(\Omega)}\;,
\end{equation}
where the constant in the right-hand side is independent of $\varepsilon$ and $d$.
\end{lem}



Prior to the proof of Lemma \ref{lemenest}  and Theorem 1 we introduce several auxiliary notations. 


\subsection{Auxiliary functions}\label{sec6_1} In this part of the proof, auxiliary functions will be introduced that will allow the remainder $\BR_N$ to be estimated. The first functions to be considered are the cut-off functions whose supports are located  in the vicinity of the boundaries of $\Omega_N$. 

Namely, the cut-off function $\chi_\varepsilon^{(k)}\in C_0^\infty(B_{3\varepsilon}^{(k)})$, $1\le k\le N$ will be used that is equal to 1 inside the ball $B_{2\varepsilon}^{(k)}$. A  cut-off function $\chi_0$ is also required and will allow for certain domains of integration to be concentrated near the boundary $\partial \Omega$. With the set $\CV_\delta=\{\Bx\in \Omega: 0<\text{dist}(\Bx, \partial \Omega)<\delta\}$ we define $\chi_0 \in C^\infty_0(\CV)$, where $\CV=\CV_{1/2}$. The function $\chi_0$ is equal to 1 on $\CV_{1/4}$, and zero when $\Bx\in \overline{\Omega}\backslash \overline{\CV}$.

Now vector functions $\BGY_k$, $k=0, 1, \dots, N$, are introduced that satisfy the conditions:
\begin{equation}\label{eqenest1}
\BGY_0(\Bx)=-\BR_N(\Bx)\;,\quad \text{ for } \Bx \in \partial \Omega\;,
\end{equation}
and 
\begin{equation}\label{eqenest2}
T_n(\nabla_\Bx)\BGY_p(\Bx)=-T_n(\nabla_\Bx)\BR_N(\Bx)\;, \quad\text{ for } \Bx \in \partial \omega_\varepsilon^{(p)}\;, p=1,\dots,N\;.
\end{equation}
Such functions will take the representations
\begin{equation}
\BGY_0(\Bx)=\sum^N_{j=1}\left\{\BQ^{(j)}_\varepsilon(\Bx)-(\BGX(\nabla_{\Bw})^T\Gamma(\Bw, \Bx))^T\BM^{(j)}_\varepsilon\Big|_{\Bw=\BO^{(j)}}\right\}\BC^{(j)}\label{eqenest3}
\end{equation}
and for $1\le k \le N$
\begin{eqnarray}
\BGY_k(\Bx)&=&\Bu(\Bx)-\BGx(\Bx-\BO^{(k)})(\BGx(\nabla_{\Bx})^T\Bu(\Bx))\Big|_{\Bx=\BO^{(k)}} -\BGX(\Bx-\BO^{(k)})(\BGX(\nabla_{\Bx})^T\Bu(\Bx))\Big|_{\Bx=\BO^{(k)}}\nonumber\\
&&-(\BD(\nabla_\Bw)^TH(\Bw, \Bx))^T\BM^{(k)}_\varepsilon\BC^{(k)}\Big|_{\Bw=\BO^{(k)}}\nonumber\\
&&+\sum_{\substack{j\ne k\\1\le j\le N}}\left\{\BQ^{(j)}_\varepsilon(\Bx)-(\BGX(\nabla_{\Bw})^TH(\Bw, \Bx))^T\BM^{(j)}_\varepsilon\Big|_{\Bw=\BO^{(j)}}\right\}\BC^{(j)}\nonumber\\
&&-\sum_{\substack{j\ne k\\1\le j\le N}}\BGX(\Bx-\BO^{(k)})\BGX(\nabla_{\Bx})^T(\BGX(\nabla_{\Bw})^TG(\Bw,\Bx))^T\BM^{(j)}_\varepsilon\BC^{(j)}\Big|_{\substack{\Bx=\BO^{(k)}\\\Bw=\BO^{(j)}}}\;.\label{eqenest4}
\end{eqnarray}
With these choices for the functions $\BGY_k$, $0\le k \le N$, it can be verified that they indeed satisfy (\ref{eqenest1}) and (\ref{eqenest2}).

Also note that for $k=1,\dots, N$ it can be checked that
\begin{equation}\label{eqenest5}
\int_{\partial \omega^{(k)}_\varepsilon}T_n(\nabla_{\Bx})\BGY_k(\Bx)dS_{\Bx}=\BO\; \quad \text{ and }\quad \int_{\partial\omega^{(k)}_\varepsilon}\BJ(\Bx-\BO^{(k)}) T_n(\nabla_\Bx)\BGY_k(\Bx)d\Bx=\BO\;.
\end{equation}
In the sequel, we also use the same notation $\BR_N$ to denote the extension of the remainder into the regions $\omega_\Gve^{(k)}$, $1\le k\le N$, similar to \cite{Stein}.

Later, the constant vectors
\begin{equation}\label{convec1}
\Br^{(k)}=\frac{1}{|B^{(k)}_{3\varepsilon}|} \int_{B^{(k)}_{3\varepsilon}}\BJ(\nabla_\Bx)\BR_N(\Bx)d\Bx\;, \quad 1\le k \le N\;,
\end{equation}
and
\begin{equation}\label{convec2}
\overline{\BR_N}^{(k)}=\frac{1}{|B_{3\varepsilon}^{(k)}|}\int_{B_{3\varepsilon}^{(k)}}(\BR_N(\Bx)+\BJ(\Bx-\BO^{(k)})\Br^{(k)})d\Bx\;\;, \quad 1\le k\le N\;,
\end{equation}
will also be required. Using these constants,  a rigid body displacement can be constructed in the form   $\BR_N(\Bx)+\BJ(\Bx)\Br^{(k)}$ that satisfies
\begin{equation}\label{normcon}
\int_{B^{(k)}_{3\varepsilon}}\BGn(\BR_N(\Bx)+\BJ(\Bx-\BO^{(k)})\Br^{(k)})d\Bx=\mathbb{O}_{3\times 3}\;,
\end{equation}
and 
\begin{equation}\label{normcon1}
\int_{B^{(k)}_{3\varepsilon}}(\BR_N(\Bx)+\BJ(\Bx-\BO^{(k)})\Br^{(k)}-\overline{\BR_N}^{(k)})d\Bx=\BO\;.
\end{equation}
\subsection{Estimate for the energy in terms of the functions $\BGY_k$}\label{sec6_2}
Here it is shown that
\begin{eqnarray}
\label{eq1enpsi}
\int_{\Omega_N}\text{tr}(\BGs(\BR_N)\Be(\BR_N))d\Bx&\le&
 \text{Const }\left\{\int_{\CV}
|\BGY_0|^2d\Bx+\int_{\CV}
|\BE(\BGY_0)|^2d\Bx+\sum_{k=1}^N\int_{B^{(k)}_{3\varepsilon}}
|\BE(\BGY_k)|^2
d\Bx \right\}\;.
\end{eqnarray}
First, set
\begin{equation}
\label{eqenest6}\BW=\BR_N+\chi_0\BGY_0\qquad \text{ and }\qquad \BU=\BR_N+\sum_{k=1}^N\chi_\varepsilon^{(k)} \BGY_k\;.
\end{equation}
Note that according to (\ref{eqenest1}) and (\ref{eqenest2}), $\BW=\BO$ for $\Bx\in \partial\Omega$ and $T_n(\nabla_\Bx)\BU=\BO$, for $\Bx\in \cup_{k=1}^N \partial \omega_\varepsilon^{(k)}$. As a result,  after applying Betti's formula, it is possible to show that 
\[\int_{\Omega_N}\text{tr}(\BGs(\BW)\Be(\BU))d\Bx=-\int_{\Omega_N}\BW\cdot L(\nabla_{\Bx})\BU d\Bx\;.\]
Recall  the supports of the cut-off functions $\chi_0$ and $\chi^{(k)}_\varepsilon$, $k=1,\dots, N$ do not intersect, and $\BR_N$ satisfies the homogeneous Lam\'e equation in $\Omega_N$. Thus  after replacing $\BU$ and $\BW$ with their definitions in (\ref{eqenest6}), the preceding identity reduces to 
\begin{equation}\label{eqenest7}
\int_{\Omega_N}\text{tr}(\BGs(\BR_N+\chi_0\BGY_0)\Be(\BR_N+\sum_{k=1}^N\chi_\varepsilon^{(k)} \BGY_k))d\Bx=-\sum_{k=1}^N  \int_{B^{(k)}_{3\varepsilon}\backslash \overline{\omega_\varepsilon^{(k)}}} \BR_N\cdot L(\nabla_\Bx)(\chi^{(k)}_\varepsilon\BGY_k)d\Bx
\end{equation}
which can be further simplified by expanding the left-hand side using the linearity of the stress and strain tensors to give the inequality
\begin{eqnarray}\label{eqenest8}
\int_{\Omega_N}\text{tr}(\BGs(\BR_N)\Be(\BR_N))d\Bx&\le &\Sigma_1+\Sigma_2+\Sigma_3
\end{eqnarray}
where
\begin{eqnarray}
&&\Sigma_1=\Big|\int_{\CV}\text{tr}(\BGs(\chi_0\BGY_0)\Be(\BR_N))d\Bx\Big|\;,
 \nonumber \\
&&\Sigma_2=\Big|\sum_{k=1}^N  \int_{B^{(k)}_{3\varepsilon}\backslash \overline{\omega_\varepsilon^{(k)}}} \BR_N\cdot L(\nabla_\Bx)(\chi^{(k)}_\varepsilon\BGY_k)d\Bx\Big|\;,\nonumber\\
&&\Sigma_3=\Big|\sum_{k=1}^N \int_{B^{(k)}_{3\varepsilon}\backslash \overline{\omega_\varepsilon^{(k)}}}\text{tr}(\BGs(\BR_N)\Be(\chi_\varepsilon^{(k)}\BGY_k))d\Bx\Big|\;.\label{Sigs}
\end{eqnarray}
Next, to derive (\ref{eq1enpsi}), $\Sigma_j$, $j=1,2,3$ are estimated.
\subsubsection{Estimate for $\Sigma_1$}
The term $\Sigma_1$, by the Cauchy inequality and the Schwarz inequality, admits the estimate
\begin{eqnarray}\label{eqenest9}
\Sigma_1&\le&\int_{\CV}[\text{tr}(\BGs(\chi_0\BGY_0)\BGs(\chi_0\BGY_0))]^{1/2}[S(\BR_N)]^{1/2}d\Bx\nonumber\\
&\le &\left(\int_{\CV}\text{tr}(\BGs(\chi_0\BGY_0)\BGs(\chi_0\BGY_0))
d\Bx\right)^{1/2}\left(\int_{\CV}
S(\BR_N)
d\Bx\right)^{1/2}\;.
\end{eqnarray}
Here, the quantity  $S(\BU)$ is defined in (\ref{eqtreE}).
Since  the inequalities 
\begin{equation}
\label{eqSest1}
\text{tr}(\BGs(\Bv)\BGs(\Bv))\le c_1S(\Bv)\quad \text{ where }\quad c_1=\left\{\begin{array}{ll}(3\lambda+2\mu)^2&\quad \text{if } 0\le \nu\le 1/2\;,\\
4\mu^2& \quad \text{if }  -1\le  \nu< 0\;,
\end{array}
\right.
\end{equation}
and
\begin{equation}
\label{imest}
\text{tr}(\BGs(\Bv)\Be(\Bv))\ge c_2S(\Bv)\quad \text{ where }\quad c_2=\left\{\begin{array}{ll}2\mu&\quad \text{if } 0\le \nu\le 1/2\;,\\
3\lambda+2\mu& \quad \text{if }  -1\le  \nu< 0\;,
\end{array}\right.\end{equation}
hold for a vector function $\Bv$, it can then be asserted from (\ref{eqSest1}), (\ref{imest}) and  (\ref{eqenest9}) that
\begin{eqnarray}\label{eqenest10}
\Sigma_1 &\le&\text{Const }\left(\int_{\CV}
S(\chi_0\BGY_0)d\Bx\right)^{1/2}\left(\int_{\Omega_N}\text{tr}(\BGs(\BR_N)\Be(\BR_N))d\Bx\right)^{1/2}.
\end{eqnarray}

\subsubsection{Estimate for $\Sigma_2$}
Note  that 
\begin{equation}\label{eqenest13}
\int_{B_{3\varepsilon}^{(k)}\backslash \overline{\omega^{(k)}_\varepsilon}}(\BJ(\Bx)\Br^{(k)}-\overline{\BR_N}^{(k)})\cdot L(\nabla_{\Bx})(\chi^{(k)}_\varepsilon\BGY_k)d\Bx=0\;,
\end{equation}
where the definitions of $\Br^{(k)}$ and $\overline{\BR_N}^{(k)}$ are found in (\ref{convec1}) and (\ref{convec2}).
Identity (\ref{eqenest13}) appears as a result of the application of  the Betti formula in $B_{3\varepsilon}^{(k)}\backslash \overline{\omega^{(k)}_\varepsilon}$ as follows:
\begin{eqnarray}\label{eqenest14}
&&\int_{B_{3\varepsilon}^{(k)}\backslash \overline{\omega^{(k)}_\varepsilon}}(\BJ(\Bx-\BO^{(k)})\Br^{(k)}-\overline{\BR_N}^{(k)})\cdot L(\nabla_{\Bx})(\chi^{(k)}_\varepsilon\BGY_k)d\Bx
=-\int_{B_{3\varepsilon}^{(k)}\backslash \overline{\omega^{(k)}_\varepsilon}}\chi^{(k)}_\varepsilon\BGY_k\cdot L(\nabla_{\Bx})(\BJ(\Bx-\BO^{(k)})\Br^{(k)}-\overline{\BR_N}^{(k)}) d\Bx\nonumber\\
&&+\int_{\partial (B_{3\varepsilon}^{(k)}\backslash \overline{\omega^{(k)}_\varepsilon)}}\{(\BJ(\Bx-\BO^{(k)})\Br^{(k)}-\overline{\BR_N}^{(k)})\cdot T_n(\nabla_{\Bx})(\chi^{(k)}_\varepsilon\BGY_k)-\chi^{(k)}_\varepsilon\BGY_k\cdot T_n(\nabla_{\Bx})(\BJ(\Bx-\BO^{(k)})\Br^{(k)}-\overline{\BR_N}^{(k)})\} ds_\Bx\;.
\nonumber\\
\end{eqnarray}
The first integral on the right is zero since all rigid body displacements are solutions of the  homogeneous  Lam\'e system. They also produce zero traction and this together with the definition of $\chi_\varepsilon^{(k)}$, $1\le k \le N$, shows that
\[\int_{B_{3\varepsilon}^{(k)}\backslash \overline{\omega^{(k)}_\varepsilon}}(\BJ(\Bx-\BO^{(k)})\Br^{(k)}-\overline{\BR_N}^{(k)})\cdot L(\nabla_{\Bx})(\chi^{(k)}_\varepsilon\BGY_k)d\Bx
=\int_{\partial  \omega^{(k)}_\varepsilon}(\BJ(\Bx-\BO^{(k)})\Br^{(k)}-\overline{\BR_N}^{(k)})\cdot T_n(\nabla_{\Bx})\BGY_kds_\Bx\;,\]
and owing to (\ref{eqenest5}) the right-hand side is zero.

In addition to (\ref{eqenest13}), the next identity is also true
\begin{equation}\label{eqenest15}
\int_{B_{3\varepsilon}^{(k)}\backslash \overline{\omega^{(k)}_\varepsilon}} (\BR_N(\Bx)+\BJ(\Bx-\BO^{(k)})\Br^{(k)}-\overline{\BR_N}^{(k)})\cdot L(\nabla_\Bx)(\chi_\varepsilon^{(k)}(\BJ(\Bx-\BO^{(k)})\BGy^{(k)}-\overline{\BGY_k}))d\Bx=0\;,
\end{equation}
where similar to (\ref{convec1}) and (\ref{convec2})
\[\BGy^{(k)}=\frac{1}{|B_{3\varepsilon}^{(k)}|} \int_{ B_{3\varepsilon}^{(k)}}\BJ(\nabla_\Bx)\BGY_k(\Bx)ds_\Bx, \quad 1 \le k\le N\;,\]
and
\[\overline{\BGY_k}=\frac{1}{|B_{3\varepsilon}^{(k)}|}\int_{B_{3\varepsilon}^{(k)}}(\BGY_k(\Bx)+\BJ(\Bx-\BO^{(k)})\BGy^{(k)})d\Bx\;\;, \quad 1\le k\le N\;.\]
Here (\ref{eqenest15}) follows from applying the  Betti formula inside $B^{(k)}_{3\varepsilon}\backslash \overline{\omega^{(k)}_\varepsilon}$ and making use of the fact that $\BR_N$ is a solution of the homogeneous Lam\'e equation in $\Omega_N$ and that it satisfies the 
 conditions (\ref{othocond}).

Therefore, the term $\Sigma_2$ in  (\ref{Sigs}), in combination with (\ref{eqenest13}) and (\ref{eqenest15}), is also written as
\begin{eqnarray}\label{eqn_N2}
\Sigma_2
&=&\Big|\sum_{k=1}^N\int_{B_{3\varepsilon}^{(k)}\backslash \overline{\omega^{(k)}_\varepsilon}} (\BR_N(\Bx)+\BJ(\Bx-\BO^{(k)})\Br^{(k)}-\overline{\BR_N^{(k)}})\cdot L(\nabla_\Bx)(\chi_\varepsilon^{(k)}(\BGY_k(\Bx)+\BJ(\Bx-\BO^{(k)})\BGy^{(k)}-\overline{\BGY_k}))d\Bx\Big|\;.\nonumber \\
\end{eqnarray}
The Schwarz inequality followed by the Cauchy inequality shows that $\Sigma_2$ is majorised by
\begin{eqnarray*}
&&\text{Const }  \left(\sum_{k=1}^N\int_{B_{3\varepsilon}^{(k)}
 } |(\BR_N(\Bx)+\BJ(\Bx-\BO^{(k)})\Br^{(k)}-\overline{\BR_N}^{(k)})|^2d\Bx\right)^{1/2}\\
&&\times  \left( \sum_{k=1}^N\int_{B_{3\varepsilon}^{(k)}
 } |L(\nabla_\Bx)(\chi_\varepsilon^{(k)}(\BGY_k(\Bx)+\BJ(\Bx-\BO^{(k)})\BGy^{(k)}-\overline{\BGY_k}))|^2d\Bx\right)^{1/2},
\end{eqnarray*}
where $\BR_N$ has been smoothly extended inside  $\omega_\Gve^{(k)}$. Then 
Poincare's inequality shows that  in $B_{3\varepsilon}^{(k)}$, $k=1,\dots, N,$
\begin{equation}\label{step1}
\left(\int_{B_{3\varepsilon}^{(k)}
} |(\BR_N(\Bx)+\BJ(\Bx-\BO^{(k)})\Br^{(k)}-\overline{\BR_N^{(k)}})|^2d\Bx\right)^{1/2}\le \text{Const }\varepsilon \left(\int_{B_{3\varepsilon}^{(k)}
} |\nabla(\BR_N(\Bx)+\BJ(\Bx-\BO^{(k)})\Br^{(k)})|^2d\Bx\right)^{1/2}\;. 
\end{equation}
Next as a result of condition (\ref{normcon}),  the Friedrichs inequality   can be used, similar to \cite{Horgan}, to give the estimate
\begin{equation}
\label{step2}
\left(\int_{B_{3\varepsilon}^{(k)}
} |\nabla(\BR_N(\Bx)+\BJ(\Bx-\BO^{(k)})\Br^{(k)})|^2d\Bx\right)^{1/2}\le \text{Const }\varepsilon \left(\int_{B_{3\varepsilon}^{(k)}
} S(\BR_N)d\Bx\right)^{1/2}\;. 
\end{equation}
This argument together with   (\ref{imest}) and (\ref{eqn_N2}) shows that
\begin{eqnarray}
\label{eqenest16}
\Sigma_2&\le &\text{Const }\varepsilon \left(\sum_{k=1}^N\int_{B_{3\varepsilon}^{(k)}
}  \text{tr}(\BGs(\BR_N)\Be(\BR_N))d\Bx\right)^{1/2}\nonumber \\
& &\times \left( \sum_{k=1}^N\int_{B_{3\varepsilon}^{(k)}
} |L(\nabla_\Bx)(\chi_\varepsilon^{(k)}(\BGY_k(\Bx)+\BJ(\Bx-\BO^{(k)})\BGy^{(k)}-\overline{\BGY_k}))|^2d\Bx\right)^{1/2}\;.
\end{eqnarray}
By computing derivatives and taking into account the definition of the cut-off functions $\chi_k$, $k=1, \dots, N$, an estimate for the second integrand on the right can be established in the form 
\begin{eqnarray}
|L(\nabla_\Bx)(\chi_\varepsilon^{(k)}(\BGY_k(\Bx)+\BJ(\Bx)\BGy^{(k)}-\overline{\BGY_k}))|^2&\le&\text{Const } \varepsilon^{-2}\Big\{|\nabla(\BGY_k(\Bx)+\BJ(\Bx-\BO^{(k)})\BGy^{(k)})|^2\nonumber\\
&&\qquad\qquad+\varepsilon^{-2}|\BGY_k(\Bx)+\BJ(\Bx-\BO^{(k)})\BGy^{(k)}-\overline{\BGY_k}|^2\Big\}
\;,\label{eqenest17}
\end{eqnarray}
where the $L(\nabla_\Bx) \BGY_k=\BO$, for $\Bx \in B_{3\varepsilon}^{(k)}$ has been used.

Thus, (\ref{eqenest17}) together with the  application of the  Poincar\'e inequality  and the Friedrichs inequality  inside $B_{3\varepsilon}^{(k)}$ leads to
\begin{eqnarray}\label{Lfnest}
&&\left( \sum_{k=1}^N\int_{B_{3\varepsilon}^{(k)}
} |L(\nabla_\Bx)(\chi_\varepsilon^{(k)}(\BGY_k(\Bx)+\BJ(\Bx-\BO^{(k)})\BGy^{(k)}-\overline{\BGY_k}))|^2d\Bx\right)^{1/2}\nonumber\\
&\le& \text{Const }\varepsilon^{-1}\left(\sum_{k=1}^N\int_{B_{3\varepsilon}^{(k)}} S(\BGY_k)d\Bx
\right)^{1/2}.
\end{eqnarray}
Combined with (\ref{eqenest16}) and  the fact that
\[\left(\sum_{k=1}^N\int_{B_{3\varepsilon}^{(k)}
}  \text{tr}(\BGs(\BR_N)\Be(\BR_N))d\Bx\right)^{1/2}\le  \left(\int_{\Omega_N}  \text{tr}(\BGs(\BR_N)\Be(\BR_N))d\Bx\right)^{1/2}\;,\]
(\ref{Lfnest}) then yields 
\begin{equation}\label{eqenest18}
\Sigma_2 \le \text{Const }\left(\int_{\Omega_N}  \text{tr}(\BGs(\BR_N)\Be(\BR_N))d\Bx\right)^{1/2}\left(\sum_{k=1}^N\int_{B_{3\varepsilon}^{(k)}} S(\BGY_k)d\Bx
\right)^{1/2}\;.
\end{equation}

\subsubsection{Estimate for $\Sigma_3$}
Owing to the Betti formula,  Lemma \ref{lemuNapp} and the assumption that  the support of the cut-off function $\chi^{(k)}_\varepsilon$ is contained in $B^{(k)}_{3\varepsilon}$, we deduce
\begin{eqnarray*}
&&\int_{B^{(k)}_{3\varepsilon}\backslash \overline{\omega_\varepsilon^{(k)}}}\text{tr}(\BGs(\BR_N)\Be(\chi_\varepsilon^{(k)}\{\BJ(\Bx-\BO^{(k)})\BGy^{(k)}-\overline{\BGY_k}\}))d\Bx\\
&=&-\int_{B^{(k)}_{3\varepsilon}\backslash \overline{\omega_\varepsilon^{(k)}}}\chi_\varepsilon^{(k)}\{\BJ(\Bx-\BO^{(k)})\BGy^{(k)}-\overline{\BGY_k}\}\cdot L(\nabla_\Bx)\BR_Nd\Bx =0\;.
\end{eqnarray*}
It then follows that 
\begin{equation}
\label{6.2.9a}
\int_{B^{(k)}_{3\varepsilon}\backslash \overline{\omega_\varepsilon^{(k)}}}\text{tr}(\BGs(\BR_N)\Be(\chi_\varepsilon^{(k)}\BGY_k))d\Bx
=
\int_{B^{(k)}_{3\varepsilon}\backslash \overline{\omega_\varepsilon^{(k)}}}\text{tr}(\BGs(\BR_N)\Be(\chi_\varepsilon^{(k)}\{\BGY_k+\BJ(\Bx-\BO^{(k)})\BGy^{(k)}-\overline{\BGY_k}\}))d\Bx
\;.\end{equation}
The symmetry of the functional on the right-hand side implies
\begin{eqnarray}
&&\int_{B^{(k)}_{3\varepsilon}\backslash \overline{\omega_\varepsilon^{(k)}}}\text{tr}(\BGs(\BR_N)\Be(\chi_\varepsilon^{(k)}\{\BGY_k-\BJ(\Bx-\BO^{(k)})\BGy^{(k)}-\overline{\BGY_k}\}))d\Bx \nonumber\\
&=&\int_{B^{(k)}_{3\varepsilon}\backslash \overline{\omega_\varepsilon^{(k)}}}\text{tr}(\BGs(\chi_\varepsilon^{(k)}\{\BGY_k+\BJ(\Bx-\BO^{(k)})\BGy^{(k)}-\overline{\BGY_k}\})\Be(\BR_N))d\Bx \;. \label{6.2.9b}
\end{eqnarray}
After applying the Cauchy and Schwarz inequalities to (\ref{6.2.9b}) and combining the result with (\ref{6.2.9a}), it can be derived that
\begin{eqnarray}
&&\int_{B^{(k)}_{3\varepsilon}\backslash \overline{\omega_\varepsilon^{(k)}}}\text{tr}(\BGs(\BR_N)\Be(\chi_\varepsilon^{(k)}\BGY_k))d\Bx \le  \left(\int_{B^{(k)}_{3\varepsilon}\backslash \overline{\omega_\varepsilon^{(k)}}}
S(\BR_N)
d\Bx\right)^{1/2}\nonumber\\
&& \times 
\left(\int_{B^{(k)}_{3\varepsilon}\backslash \overline{\omega_\varepsilon^{(k)}}}\text{tr}(\BGs(\chi_\varepsilon^{(k)}\{\BGY_k+\BJ(\Bx-\BO^{(k)})\BGy^{(k)}-\overline{\BGY_k}\})\BGs(\chi_\varepsilon^{(k)}\{\BGY_k+\BJ(\Bx-\BO^{(k)})\BGy^{(k)}-\overline{\BGY_k}\}))
d\Bx\right)^{1/2}
\end{eqnarray}
where $S(\BU)$ is given in (\ref{eqtreE}). Then (\ref{eqSest1}) and (\ref{imest}) provide
\begin{eqnarray}\label{eqenest11a}\int_{B^{(k)}_{3\varepsilon}\backslash \overline{\omega_\varepsilon^{(k)}}}\text{tr}(\BGs(\BR_N)\Be(\chi_\varepsilon^{(k)}\BGY_k))d\Bx &\le& \text{Const } \left(\int_{B^{(k)}_{3\varepsilon}}
S(\chi_\varepsilon^{(k)}\{\BGY_k+\BJ(\Bx-\BO^{(k)})\BGy^{(k)}-\overline{\BGY_k}\})
d\Bx\right)^{1/2}\nonumber \\ 
&&\times \left(\int_{B^{(k)}_{3\varepsilon}\backslash \overline{\omega_\varepsilon^{(k)}}}\text{tr}(\BGs(\BR_N)\Be(\BR_N))d\Bx\right)^{1/2} \;.\end{eqnarray}
Here, as a result of the inequality
\[S(u\Bv)\le \text{Const}\{|\nabla u|^2|\Bv|^2+u^2S(\Bv)\}\]
for any vector function $\Bv$ and scalar function $u$, it can be asserted that 
\begin{eqnarray*}
&&\int_{B^{(k)}_{3\varepsilon}}
S(\chi_\varepsilon^{(k)}\{\BGY_k+\BJ(\Bx-\BO^{(k)})\BGy^{(k)}-\overline{\BGY_k}\})
d\Bx\\
&\le& \text{Const} \Big\{\varepsilon^{-2}\int_{B^{(k)}_{3\varepsilon}}|\BGY_k+\BJ(\Bx-\BO^{(k)})\BGy^{(k)}-\overline{\BGY_k}|^2d\Bx+\int_{B^{(k)}_{3\varepsilon}}S(\BGY_k)d\Bx\Big\}\;.
\end{eqnarray*}
Again applying the Poincar\'e inequality and  the Friedrichs inequality in $B^{(k)}_{3 \varepsilon}$ to the first integral on the above right-hand side (similar to (\ref{step1}) and (\ref{step2})) gives
\[ \int_{B^{(k)}_{3\varepsilon}}
S(\chi_\varepsilon^{(k)}\{\BGY_k+\BJ(\Bx-\BO^{(k)})\BGy^{(k)}-\overline{\BGY_k}\})
d\Bx \le \text{Const} \int_{B^{(k)}_{3\varepsilon}}S(\BGY_k)d\Bx\;.\]
This estimate together with (\ref{eqenest11a}) yields
\begin{eqnarray}\label{eqenest11}
\Sigma_3&\le& \text{Const } \left(\int_{B^{(k)}_{3\varepsilon}}
S(\BGY_k)
d\Bx\right)^{1/2} \left(\int_{B^{(k)}_{3\varepsilon}\backslash \overline{\omega_\varepsilon^{(k)}}}\text{tr}(\BGs(\BR_N)\Be(\BR_N))d\Bx\right)^{1/2} \;.\end{eqnarray}
\subsubsection{Proof of (\ref{eq1enpsi})}
Therefore (\ref{eqenest8}), (\ref{Sigs}), (\ref{eqenest10}), (\ref{eqenest18}) and (\ref{eqenest11}) assert that 
\begin{eqnarray}\label{eqff}
\int_{\Omega_N}\text{tr}(\BGs(\BR_N)\Be(\BR_N))d\Bx&\le& \text{Const }\left\{\int_{\CV}
S(\chi_0\BGY_0)d\Bx+\sum_{k=1}^N\int_{B^{(k)}_{3\varepsilon}}
S(\BGY_k)
d\Bx \right\}\;.
\end{eqnarray}
As a result of (\ref{eqtreE}), for a vector function $\Bv$
\[S(\Bv)\le \text{Const } |\BE(\Bv)|^2\;,\]
and  this with the definition of $\chi_0$ and (\ref{eqff}) yields  (\ref{eq1enpsi}).


\subsection{Proof of Lemma \ref{lemenest} and Theorem 1. Estimation of the energy for $\BR_N$}\label{sec6_3}
The inequality (\ref{eq1enpsi}) 
leads to 
\begin{equation}\label{energyest_alpha}
\int_{\Omega_N}\text{tr}(\BGs(\BR_N)\Be(\BR_N))d\Bx\le \text{Const }\left\{
\CK+\CL+\CM+\CN\right\}\;,
\end{equation}
where
\begin{eqnarray}
&&\CK=\int_{\CV}
|\BGY_0|^2d\Bx+\int_{\CV}
|\BE(\BGY_0)|^2d\Bx\;,\\
\nonumber\\&&
\CL=\sum_{k=1}^N\int_{B^{(k)}_{3\varepsilon}}|\BE[\Bu(\Bx) -\BGX(\Bx-\BO^{(k)})(\BGX(\nabla_{\Bx})^T\Bu(\Bx))\Big|_{\Bx=\BO^{(k)}}]|^2d\Bx\;,\\
\nonumber\\&&
\CM=\sum_{k=1}^N\int_{B_{3\varepsilon}^{(k)}}\Big|\BE\Big(\sum_{\substack{j\ne k\\1\le j\le N}}\left\{\BQ^{(j)}_\varepsilon(\Bx)-(\BGX(\nabla_{\Bw})^TH(\Bw, \Bx))^T\BM^{(j)}_\varepsilon\Big|_{\Bw=\BO^{(j)}}\right\}\BC^{(j)}\nonumber\\
&&\qquad\qquad-\sum_{\substack{j\ne k\\1\le j\le N}}\BGX(\Bx-\BO^{(k)})\BGX(\nabla_{\Bx})^T(\BGX(\nabla_{\Bw})^TG(\Bw,\Bx))^T\BM^{(j)}_\varepsilon\BC^{(j)}\Big|_{\substack{\Bx=\BO^{(k)}\\\Bw=\BO^{(j)}}}\Big)\Big|^2d\Bx\;,\\
\nonumber\\&&\CN=\sum_{k=1}^N\int_{B^{(k)}_{3\varepsilon}}\Big|\BE\left((\BD(\nabla_\Bw)^TH(\Bw, \Bx))^T\BM^{(k)}_\varepsilon\BC^{(k)}\Big|_{\Bw=\BO^{(k)}}
\right)\Big|^2d\Bx\;.
\end{eqnarray}

Owing to the representation of $\BGY_0$ in (\ref{eqenest3}) and  Lemma \ref{lemQ}, the term $\CK$ admits the estimate
\begin{eqnarray*}
\CK&\le &\text{Const }\varepsilon^8 \Big\{\int_\CV\Big(\sum_{j=1}^N\frac{|\BC^{(j)}|}{|\Bx-\BO^{(j)}|^3}\Big)^2d\Bx+\int_\CV\Big(\sum_{j=1}^N\frac{|\BC^{(j)}|}{|\Bx-\BO^{(j)}|^4}\Big)^2d\Bx\Big\}\\
&\le & \text{Const }\varepsilon^8 \sum_{j=1}^N|\BC^{(j)}|^2\sum_{j=1}^N \Big\{\int_\CV\frac{1}{|\Bx-\BO^{(j)}|^6}d\Bx+\int_\CV\frac{1}{|\Bx-\BO^{(j)}|^8}d\Bx\Big\}\;,
\end{eqnarray*}
where the last estimate has been obtained through the Cauchy inequality. Since $\text{dist}(\partial \Omega, \partial \omega)\ge 1$, the final estimate for $\CK$, after applying Lemma \ref{lem_invert}, is 
\begin{equation}\label{estK}
\CK\le \text{Const }\varepsilon^8d^{-3}\sum_{j=1}^N|\BC^{(j)}|^2\le\text{Const }\varepsilon^8d^{-6}\|\BE(\Bu)\|^2_{L_\infty(\Omega)} \;.
\end{equation}
To estimate $\CL$, the Taylor approximation is used to expand the first-order derivatives of the function $\Bu$ about $\Bx=\BO^{(k)}$, as follows
\begin{eqnarray*}
\CL&= &\sum_{k=1}^N\int_{B^{(k)}_{3\varepsilon}}|\BE[\Bu(\Bx)] -\BE[\Bu(\Bx)]\Big|_{\Bx=\BO^{(k)}}]|^2d\Bx\\
&\le &\text{Const }\varepsilon^5\sum_{p=1}^N\|\nabla \otimes \BE[\Bu(\Bx))]\Big|_{\Bx=\BO^{(k)}}\|^2
\;.
\end{eqnarray*}
A local regularity estimate for the second-order derivatives of the components of $\Bu$  inside $\omega$ then leads to 
\begin{equation}\label{estL}
\CL\le 
 \text{Const }\varepsilon^5d^{-3}\|\BE(\Bu)\|^2_{L_\infty(\Omega)}\;.
\end{equation}

By using the boundary condition for the regular part $H$ (see  Section 3), the term $\CM$ can be written in the form
\begin{eqnarray}
\CM&=&\sum_{k=1}^N\int_{B_{3\varepsilon}^{(k)}}\Big|\sum_{\substack{j\ne k\\1\le j\le N}}\Big[\BE\left(\BQ^{(j)}_\varepsilon(\Bx)-(\BGX(\nabla_{\Bw})^T\Gamma(\Bw, \Bx))^T\Big|_{\Bw=\BO^{(j)}}\right)\nonumber\\
&&-\BE\left((\BGX(\nabla_{\Bw})^TG(\Bw,\Bx))^T\Big|_{ \Bw=\BO^{(j)}}\right)\Big|_{\Bx=\BO^{(k)}}\Big]\BM^{(j)}_\varepsilon\BC^{(j)}\Big|^2d\Bx\;.\nonumber
\end{eqnarray}
Next, using Lemma \ref{lemQ} 
 and the Taylor expansion about $\Bx=\BO^{(k)}$ of the second-order derivatives of the components of $G$, establishes the estimate
\begin{eqnarray}
\CM&\le &\text{Const }\varepsilon^{11}\sum_{k=1}^N\Big|\sum_{\substack{j\ne k\\1\le j\le N}}\frac{|\BC^{(j)}|}{|\BO^{(k)}-\BO^{(j)}|^4}\Big|^2d\Bx\nonumber\\
&\le &\text{Const }\varepsilon^{11}\sum^N_{p=1}|\BC^{(p)}|^2\sum_{k=1}^N\sum_{\substack{j\ne k\\1\le j\le N}}\frac{1}{|\BO^{(k)}-\BO^{(j)}|^8}\;.
\end{eqnarray}
Lemma \ref{lem_invert} then yields the final estimate for $\CM$:
\begin{eqnarray}
\CM
&\le &\text{Const }\varepsilon^{11}d^{-6}\sum^N_{p=1}|\BC^{(p)}|^2\iint_{\substack{\omega\times \omega:\\ |\Bx-\By|\ge d }}\frac{d\Bx d\By}{|\Bx-\By|^8}\nonumber\\
&\le &\text{Const }\varepsilon^{11}d^{-8}\sum^N_{p=1}|\BC^{(p)}|^2\le \text{Const }\varepsilon^{11}d^{-11}\|\BE(\Bu)\|^2_{L_\infty(\Omega)}\label{estM}\;.
\end{eqnarray}
Since the derivatives of the components of $H$ are bounded within the cloud $\omega$, we deduce
\begin{equation}\label{estN}
\CN\le \text{Const } \varepsilon^9\sum_{k=1}^N|\BC^{(k)}|^2 \le \text{Const } \varepsilon^9 d^{-3}            \|\BE(\Bu)\|^2_{L_\infty(\Omega)}\;.
\end{equation}
The energy estimate contained in Lemma \ref{lemenest} is then proved by combining (\ref{estK}), (\ref{estL}), (\ref{estM}), (\ref{estN}) and (\ref{energyest_alpha}).
\hfill $\Box$

Now we prove Theorem \ref{th1}. It remains to consider the formal approximation for $\Bu_N$ in Lemma \ref{lemuNapp}, which relies on the solvability of a particular algebraic system (\ref{alg_sys_eq}). The solvability of this system was proved in Lemma \ref{lem_invert}, which together with the energy estimate in Lemma \ref{lemenest}, proves Theorem \ref{th1}.
\hfill $\Box$

\section{Illustration: Simplified asymptotic formulae}\label{simp_asy}
In this section, we present simplified asymptotic formulae for $\Bu_N$ in the far-field region away from the cloud of voids and also in the case when an infinite elastic medium containing the cloud is considered. It is also shown in Appendix C that for spherical voids, the model boundary layers of Problem 3  of Section \ref{mod_field} can be constructed explicitly in the closed form, along with the dipole matrices for these spherical cavities.

\subsection{Far-field approximation to $\Bu_N$}
Given the dipole matrices $\BM^{(k)}_\varepsilon$, $1\le k\le N$, the asymptotic formula (\ref{uNapprox}) of Theorem \ref{th1} is simplified under the constraint that the point of measurement of the displacement is distant from the cloud of voids.

\begin{coro}\label{lem_simp}Let $\text{\emph{dist}}(\Bx, \omega)>1$. The asymptotic formula for $\Bu_N$ admits the form
\begin{equation}\label{eq_simp}
\Bu_N(\Bx)=\Bu(\Bx)+\sum_{k=1}^N(\BGX(\nabla_\Bz)^TG(\Bz, \Bx))^T\BM_\varepsilon^{(k)}\Big|_{\Bz=\BO^{(k)}}\BC^{(k)}+\BF_N(\Bx)\;,
\end{equation}
where $\BC^{(k)}$, $k=1,\dots, N,$ satisfy the system $(\ref{alg_sys_eq})$, 
\[\BF_N(\Bx)=O\left(\sum_{k=1}^N\frac{\varepsilon^4|\BC^{(k)}|}{|\Bx-\BO^{(k)}|^4}\right)+\BR_N\;,\]
and $\BR_N$ satisfies $(\ref{eqestRN})$.
\end{coro}

\emph{Proof. }Formula (\ref{eq_simp}) follows from Lemma \ref{lemQ}. \hfill $\Box$

\vspace{0.1in}It is noted that in the simplified representation (\ref{eq_simp}) for $\Bu_N$, information about the small voids is contained in their dipole characteristics represented by $\BM^{(k)}_\varepsilon$, $1\le k \le N$. 
In particular, if the voids are spherical cavities of radius $a^{(k)}_\varepsilon$ with centre $\BO^{(k)}$, $1\le k \le N$, then the dipole matrix is given by
\begin{equation}\label{Meps1}\begin{array}{c}
\BM^{(k)}_\varepsilon=\displaystyle{-\frac{(\lambda+2\mu)\pi (a^{(k)}_\varepsilon)^3}{\mu(9\lambda+14\mu)}}\left[\begin{array}{ccc}
\BCM^{(1)}&& \mathbb{O}_{3\times 3}\\
\mathbb{O}_{3\times 3}&& \BCM^{(2)}\end{array}\right]\end{array}\end{equation}
with
\begin{equation}\label{Meps2}
\begin{array}{c}
\BCM^{(1)}=\displaystyle{\left[\begin{array}{ccccc}
m&& m-40\mu^2&& m-40\mu^2\\
m-40\mu^2&& m&& m-40\mu^2\\
m-40\mu^2&& m-40\mu^2&& m
\end{array}
\right]\;,\qquad m=9\lambda^2+20\lambda\mu+36\mu^2\;,} \\ \\
\BCM^{(2)}=40\mu^2\mathbb{I}_{3\times 3}\;.
\end{array}
\end{equation}
It is noted that the matrix $\BM^{(k)}_\varepsilon$ for the spherical cavity in the infinite space is negative definite.
Thus (\ref{Meps1}), (\ref{Meps2}), together with Corollary \ref{lem_simp} gives the far-field approximation for $\Bu_N$ in an elastic solid  containing a cloud of arbitrary spherical cavities.
\subsection{Far-field approximation for $\Bu_N$ in an infinite elastic medium with a cloud of voids}
Here we consider the problem when $\Omega=\mathbb{R}^3$, so that $\Omega_N=\mathbb{R}^3\backslash \overline{\cup_{j=1}^N \omega^{(j)}_\varepsilon}$ is the infinite space containing a cloud of voids.

In this scenario, we search for the approximation to $\Bu_N$ which is now a solution of the problem:
\begin{equation}\label{eq_UNR_1}
L(\nabla_\Bx)\Bu_N(\Bx)+\Bf(\Bx)=\BO\;, \quad \Bx\in \Omega_N\;,
\end{equation}
\begin{equation}\label{eq_UNR_2}
T_n(\nabla_\Bx)\Bu_N(\Bx)=\BO\;,\quad  \Bx\in \partial \omega^{(j)}_\varepsilon\;, 1\le j \le N\;,
\end{equation}
\begin{equation}\label{decay_UNR}
\Bu_\BN(\Bx)=O(|\Bx|^{-2})\;, \quad \text{ for } \quad |\Bx|\to \infty\;.
\end{equation}
The vector function $\Bf$ is also supplied with the conditions that
\[\int_{\Omega_N}\Bf(\Bx)d\Bx=\BO\;, \qquad \int_{\Omega_N}\Bx \times \Bf(\Bx)d\Bx=\BO\;,\]
and the support of $\Bf$, as before, is chosen to satisfy $\text{dist}(\partial \omega, \text{supp }\Bf)=O(1)$.

Finally, before stating results concerning the approximation of $\Bu_N$, we further introduce some model quantities. We  require the field $\Bu$ which solves the problem (\ref{up1}) and (\ref{upbc}), and that is  also supplied with the additional condition of decay at infinity (\ref{decay_UNR}). The matrix 
\[\BP=\left\{\begin{array}{ll}
\BGX(\nabla_{\Bx})^T(\BGX(\nabla_\By)^T\Gamma(\By, \Bx))^T\Big|_{\substack{\Bx=\BO^{(i)}\\ \By=\BO^{(j)}}}\;,&\quad \text{ if }\quad i\ne j\;,\\
\mathbb{O}_{6\times 6}\;,&\quad \text{otherwise}\;,
\end{array}\right.\]
is also needed in the next result. We note that in the considered case the regular part $H\equiv 0$, so that Green's tensor in $\Omega$ is $G\equiv\Gamma$, the Kelvin-Somigliana tensor, which is defined in (\ref{Gamma_matrix}).

First, as a direct consequence of Corollary \ref{lem_simp} we have 
\begin{coro}\label{lem_simp_1}Let $\text{dist}(\Bx, \omega)>1$, then the asymptotic formula for $\Bu_N$ admits the form
\begin{equation}\label{eq_simp_1}
\Bu_N(\Bx)=\Bu(\Bx)+\sum_{k=1}^N(\BGX(\nabla_\Bz)^T\Gamma(\Bz, \Bx))^T\BM_\varepsilon^{(k)}\Big|_{\Bz=\BO^{(k)}}\BC^{(k)}+\mathfrak{R}_N(\Bx)\;,
\end{equation}
where 
\[\mathfrak{R}_N(\Bx)=O\left(\sum_{k=1}^N\frac{\varepsilon^4|\BC^{(k)}|}{|\Bx-\BO^{(k)}|^4}\right)+\BR_N\;,\]
$\BC=((\BC^{(1)})^T, \dots, (\BC^{(N)})^T)^T$ solves the linear  algebraic system
\begin{equation}\label{coroalg_sys_eq}
-\BV=(\mathbb{I}_{6N\times 6N}+\BP\BM)\BC\;,
\end{equation}
and $\BR_N$ satisfies $(\ref{eqestRN})$.
\end{coro}

Once again, the dipole matrix  for a spherical cavity  (see (\ref{Meps1}), (\ref{Meps2})) can be used with (\ref{eq_simp_1}) to describe the far-field behaviour of $\Bu_N$  in an infinite elastic space containing a cloud of spherical cavities.

\subsection{Uniform approximation for $\Bu_N$ in the infinite elastic space containing a cloud of voids}
Corollary \ref{lem_simp_1} can be extended to a uniform approximation $\Bu_N$, satisfying (\ref{eq_UNR_1}), (\ref{decay_UNR}), inside $\Omega_N=\mathbb{R}^3\backslash \overline{\cup_{j=1}^N \omega^{(j)}_\varepsilon}$:

\begin{coro}\label{coroth1}
Let the small parameters $\varepsilon$ and $d$ satisfy the inequality
\begin{equation}\label{coroconstraint1}
\varepsilon < c\, d\;,
\end{equation}
where $c$ is a sufficiently small constant. Then the approximation for $\Bu_N$ is given by 
\begin{equation}\label{corouNapprox}
\Bu_N(\Bx)=\Bu(\Bx)+\sum_{k=1}^N\BQ^{(k)}_\varepsilon(\Bx)\BC^{(k)}+\BR_N(\Bx)\;,
\end{equation}
and $\BR_N$ satisfies $(\ref{eqestRN})$.
\end{coro}

Matrices such as $\BQ^{(k)}_\varepsilon$ can be constructed in the explicit closed form for certain geometries. For spherical voids, the representation of this matrix is given in Appendix C. 
Thus, if the cloud $\omega$ is composed of a non-periodic arrangement of spherical voids $\omega_\varepsilon^{(j)}$, $1\le j \le N$, then the approximation stated in the previous Corollary, together with the representation of the matrix $\BQ^{(j)}_\varepsilon$ in Appendix C is readily applicable here.

\section{Concluding remarks}\label{conclusions}

A uniform asymptotic representation for a solution of a mixed boundary value problem of elasticity has been constructed and justified for a solid containing a  
 cloud of many voids. This extends significantly the results of the papers \cite{Maz_MN, Maz_MMS, MMN_Mesoelast} on meso-scale asymptotic approximations of fields in domains with multiple defects.
It is worth noting that the asymptotic representation (\ref{uNapprox}) of Theorem 1, contains an  important information about the dipole fields of a meso-scale cloud of voids.
In addition to the sum of individual contributions from the dipole fields of small voids, we have also obtained a term characterising a mutual interaction between the voids, which is often neglected in the dilute approximation procedures.
This result is significant in the area of applications linked to non-destructive testing of porous solids, where a position of a cloud and its composition can be identified through the use of the asymptotic formula (\ref{uNapprox}) accompanied by the boundary measurements for different test loading conditions.  

It is also essential to note that the meso-scale approximation (\ref{uNapprox}) is valid for different shapes of small voids  when $\varepsilon < \mbox{Const} ~ d$, for a sufficiently small constant, and this 
surpasses the range of applicability of the homogenisation approximations.

\section*{Appendix A: Local regularity of solutions to the homogeneous  Lam\'e system}
\renewcommand{\theequation}{A.\arabic{equation}}
\setcounter{equation}{0}

Here, a result concerning the estimate for the derivatives of the solution to the homogeneous Lam\'e system via their anti-derivatives is derived:

\begin{lem}\label{lem_local_reg}
Let $\Bw=\{w_i\}_{i=1}^3$ be a solution of the homogeneous Lam\'e system in a domain $\Omega$ and let $B_R\subset\Omega$, with $B_R=\{\Bx: |\Bx|<R\}$, then the estimate
\begin{equation}\label{loc_reg}
\Big| \frac{\partial w_i}{\partial x_k}(\BO) \Big| \le \text{\emph{Const }}R^{-1} \sup_{B_R}|\Bw| 
\end{equation}
holds.
\end{lem}

The proof of the last estimate uses the mean value theorem for vector functions satisfying the homogeneous Lam\'e system, as discussed below and in \cite{BramPayne}.

 \begin{lem}\label{lem_mvt}
Let $\Bw=\{w_i\}_{i=1}^3$ be a solution of the homogeneous Lam\'e system in a domain $\Omega$ and $B_R\subset\Omega$, with $B_R=\{\Bx: |\Bx|<R\}$, then:
\begin{enumerate}[(i)]
\item \begin{equation}\label{mvt_1}
w_i(\BO)=\frac{15(\lambda+\mu)}{8\pi R^4 (\lambda+4\mu)} \int_{\partial B_R} x_ix_j w_j(\Bx)ds_\Bx-\frac{3(\lambda-\mu)}{8\pi R^2(\lambda+4\mu)}\int_{\partial B_R}w_i(\Bx)ds_\Bx\;,
\end{equation}
\item  \begin{equation}\label{mvt_2}
w_i(\BO)=\frac{75(\lambda+\mu)}{8\pi R^5 (\lambda+4\mu)} \int_{ B_R} x_ix_j w_j(\Bx)d\Bx-\frac{15(\lambda-\mu)}{8\pi R^5(\lambda+4\mu)}\int_{ B_R}|\Bx|^2w_i(\Bx)d\Bx\;.
\end{equation}
\end{enumerate}
\end{lem}

\vspace{0.1in}\emph{Proof.} (i) The mean value theorem of (\ref{mvt_1}) was proved in \cite{BramPayne}. 
(ii) To derive (\ref{mvt_2}), apply  (\ref{mvt_1}) inside the ball $B_r\subset \Omega$. Then multiplying  through the resulting equation   by $r^4$ and integrating both sides with respect to $r$ between zero and $R$ yields (\ref{mvt_2}).\hfill $\Box$

\vspace{0.1in}\emph{Proof of $(\ref{loc_reg})$}:  The mean value theorem (\ref{mvt_2}) is applied in $B_R$ to the function $\displaystyle{\frac{\partial w_i}{\partial x_k}}$ as follows:
\begin{equation}\label{mvt_2_a}
\frac{\partial w_i}{\partial x_k}(\BO)=\frac{75(\lambda+\mu)}{8\pi R^5 (\lambda+4\mu)} \int_{ B_R} x_ix_j \frac{\partial w_j}{\partial x_k}(\Bx)d\Bx-\frac{15(\lambda-\mu)}{8\pi R^5(\lambda+4\mu)}\int_{ B_R}|\Bx|^2\frac{\partial w_i}{\partial x_k}(\Bx)d\Bx\;.
\end{equation}
Integration by parts then yields the two identities:
\begin{equation}\label{id_ibp_1}
 \int_{ B_R} x_ix_j \frac{\partial w_j}{\partial x_k}(\Bx)d\Bx=-\int_{B_R}(\delta_{ik}x_jw_j+x_iw_k )d\Bx+\int_{\partial B_R}n_k x_ix_jw_j ds_{\Bx}\;,
\end{equation}
\begin{equation}\label{id_ibp_2}
 \int_{ B_R} |\Bx|^2 \frac{\partial w_j}{\partial x_k}(\Bx)d\Bx=-2\int_{B_R}x_kw_jd\Bx+\int_{\partial B_R}n_k |\Bx|^2w_j ds_{\Bx}\;.
\end{equation}
Then (\ref{id_ibp_1}) and (\ref{id_ibp_2})  give the estimates:
\begin{equation}
\Big|\int_{ B_R} x_ix_j \frac{\partial w_j}{\partial x_k}(\Bx)d\Bx \Big|\le \text{Const }R^4\,\sup_{B_R}|\Bw| \quad \text{ and }\quad  \Big|\int_{ B_R} |\Bx|^2 \frac{\partial w_j}{\partial x_k}(\Bx)d\Bx\Big|\le \text{Const }R^4 \,\sup_{B_R}|\Bw|\;,
\end{equation}
and combining these with (\ref{mvt_2_a}) yields the local regularity estimate (\ref{loc_reg}). The proof is complete. \hfill $\Box$

\section*{Appendix B: Proof of (\ref{QF})
}
\renewcommand{\theequation}{B.\arabic{equation}}
\setcounter{equation}{0}
\renewcommand{\thesubsection}{B.\arabic{subsection}}
\setcounter{subsection}{0}

Here,  the proof of (\ref{QF}) is carried out by first developing an identity which will lead to an integral representation of (\ref{SP_1}) in Section \ref{sec9.1}. Then we prove some auxiliary integral identities in Section \ref{sec9.2} that are used to complete the proof of (\ref{QF}) in Section \ref{sec9.3}. 

\subsection{Poisson-type representation of the second-order derivatives of Green's tensor}\label{sec9.1}
The proof of the next lemma uses the mean value theorem for solutions of the homogeneous Lam\'e system inside disjoint balls denoted by 
$B^{(j)}=\{\Bx:|\Bx-\BO^{(j)}|<d/4\}$, $j=1,\dots, N$.

According to \cite{BramPayne} and Lemma 10 of \cite{MMN_Mesoelast}, the next result holds:

\begin{lem}For $j\ne k$, $1\le j,k\le N$, the identity 
\begin{equation}\label{lem_DDG}
\Xi_{sp}(\nabla_{\BZ})\Xi_{hq}(\nabla_{\BW}) G_{sh}(\BZ,\BW)\Big|_{\substack{\BZ=\BO^{(j)}\\ \BW=\BO^{(k)}}}=\frac{1}{(8\pi)^2 (\frac{d}{4})^6 (\lambda+4\mu)^2} \CA_{pq}^{(j,k)} 
\end{equation}
is valid, where
\begin{eqnarray}
\CA_{pq}^{(j,k)}&=&36^2 (\lambda+4\mu)^2 \CJ^{(1,j,k)}_{pq}+90(\lambda+4\mu)(\lambda+\mu)[\CJ^{(2,j,k)}_{pq}\nonumber\\ 
&&+
\CJ^{(2,k,j)}_{qp}] -18(\lambda+4\mu)(\lambda-\mu)[\CJ^{(3,j,k)}_{pq}+\CJ^{(3,k,j)}_{qp}]+225(\lambda+\mu)^2 \CJ^{(4,j,k)}_{pq}\nonumber\\
\\&&-45(\lambda^2-\mu^2)[\CJ^{(5,j,k)}_{pq}+\CJ^{(5,k,j)}_{qp}]+9(\lambda-\mu)^2\CJ^{(6,j,k)}_{pq}\;,\label{eqApq}
\end{eqnarray}
and the terms $\CJ^{(s,j,k)}_{pq}$ for $1\le s \le 6$ are 
\begin{eqnarray*}
&& \CJ^{(1,j,k)}_{pq}=\int_{B^{(j)}}\int_{B^{(k)}} \Xi_{ap}(\nabla_{\BZ})\Xi_{bq}(\nabla_{\BW})G_{ab}(\BZ,\BW)\, d\BW d\BZ\;,\\
&& \CJ^{(2,j,k)}_{pq}=\int_{B^{(j)}}\int_{B^{(k)}}  (\BZ-\BO^{(j)})_t \frac{\partial}{\partial Z_a}(\Xi_{ap}(\nabla_{\BZ}) \Xi_{bq}(\nabla_{\BW}) G_{tb}(\BZ,\BW))\, d\BW d\BZ\;,\\
&&\CJ^{(3,j,k)}_{pq}=\int_{B^{(j)}}\int_{B^{(k)}}  (\BZ-\BO^{(j)})_s \frac{\partial}{\partial Z_s} (\Xi_{ap}(\nabla_{\BZ})\Xi_{bq}(\nabla_{\BW})G_{ab}(\BZ,\BW))\,d\BW d\BZ\;,\\
&&\CJ^{(4,j,k)}_{pq}=\int_{B^{(j)}}\int_{B^{(k)}} (\BZ-\BO^{(j)})_t(\BW-\BO^{(k)})_s \frac{\partial^2}{\partial Z_a\partial W_b}(\Xi_{ap}(\nabla_{\BZ}) \Xi_{bq}(\nabla_\BW)G_{ts}(\BZ,\BW))\,d\BW d\BZ\;,\\
&&\CJ^{(5,j,k)}_{pq}=\int_{B^{(j)}}\int_{B^{(k)}} (\BZ-\BO^{(j)})_t(\BW-\BO^{(k)})_s \frac{\partial^2}{\partial Z_a \partial W_s}(\Xi_{mp}(\nabla_{\BZ})\Xi_{nq}(\nabla_{\BW}) G_{tb}(\BZ,\BW))\,d\BW d\BZ\;,\\
&& \CJ^{(6,j,k)}_{pq}=\int_{B^{(j)}}\int_{B^{(k)}} (\BZ-\BO^{(j)})_s(\BW-\BO^{(k)})_t \frac{\partial^2}{\partial Z_s \partial W_t}(\Xi_{ap}(\nabla_{\BZ})\Xi_{bq}(\nabla_{\BW}) G_{ab}(\BZ,\BW))\, d\BW d\BZ\;.
\end{eqnarray*}
\end{lem}

Before presenting the proof, it is noted that (\ref{lem_DDG}) is also a connected with the classical results of \cite{ADNI, ADNII} on estimates for solutions of elliptic partial differential equations  (eg. see Theorems 7.3 in \cite{ADNI} and Theorem 9.3 in \cite{ADNII}).

\emph{Proof. } Using the Kronecker delta the term in the left-hand side of (\ref{lem_DDG}) is rewritten as 
\begin{equation}
\label{new_DDG}
\delta_{ms}\delta_{hn} \Xi_{mp}(\nabla_{\BZ})\Xi_{nq}(\nabla_{\BW})G_{sh}(\BZ,\BW)\Big|_{\substack{\BZ=\BO^{(j)}\\ \BW=\BO^{(k)}}}\;. 
\end{equation}
From here, the term $\Xi_{mp}(\nabla_{\BZ})\Xi_{nq}(\nabla_{\BW})G_{sh}(\BZ,\BW)\Big|_{\substack{\BZ=\BO^{(j)}\\ \BW=\BO^{(k)}}}$ may be considered as entries of the matrix
\[\Xi_{mp}(\nabla_{\BZ})\Xi_{nq}(\nabla_{\BW})G(\BZ,\BW)\]
which  satisfies the homogeneous Lam\'e equation 
for $\BZ\in B^{(j)}$. The transpose of the preceding matrix also satisfies the homogenous Lam\'e equation 
for $\BW\in B^{(k)}$, $k\ne j$. Then repeating the steps of the proof of Lemma 10 in \cite{MMN_Mesoelast}, for the last matrix and inserting the resulting expression in (\ref{new_DDG}) we arrive at $\CA_{pq}^{(j,k)}$ (see (\ref{eqApq})) and the relation (\ref{lem_DDG}). The proof is complete. \hfill$\Box$

\subsection{Auxiliary integral identities}\label{sec9.2}

Now the Poisson-type representations for the second-order derivatives of Green's tensor are in place, further identities are now derived which are used in the proof Lemma \ref{lemQFest}, in the next section. From  here, we will also make use of the vector and matrix functions
\begin{equation}\label{Xi}
\BGF(\Bx)=\left\{\begin{array}{ll} \BM^{(j)}_\varepsilon \BC^{(j)}\;, &\quad \text{ if }\Bx \in \overline{B^{(j)}\;,}\\
\BO\;, &\quad \text{otherwise\;,} 
\end{array}\right.
\end{equation}
and 
\begin{equation}\label{Xi_lin}
\BGT(\Bx)=\left\{\begin{array}{ll} (\BM^{(j)}_\varepsilon \BC^{(j)})\otimes (\Bx-\BO^{(j)})\;, &\quad \text{ if }\Bx \in \overline{B^{(j)}\;,}\\
\BO\;, &\quad \text{otherwise,} 
\end{array}\right.
\end{equation}
respectively.

\begin{lem} The identities 
\begin{equation}\label{zero_id_1}
\int_{\Omega} \Theta_{mt}(\BZ)\frac{\partial}{\partial Z_n}(\Xi_{ap}(\nabla_{\BZ})G_{tb}(\BZ,\BW))\,d\BZ=0\;,
\end{equation}
\begin{equation}\label{zero_id_2}
\int_{\Omega} \GF_p(\BZ)\Xi_{ap}(\BZ)G_{ba}(\BW,\BZ)\,d\BZ=0\;,
\end{equation}
hold.\label{lem_aux_1}
\end{lem}
\emph{Proof. } We prove (\ref{zero_id_1}) and note that 
the identity (\ref{zero_id_2}) is proved in a similar way with obvious modifications.  Set 
\[f(\BW)=\int_\Omega \frac{\partial}{\partial Z_n}\Xi_{ap}(\nabla_{\BZ})G(\BW, \BZ)\BGT^T(\BZ)d\BZ\;,\]
which is the same as the left-hand side in (\ref{zero_id_1}), with the assumption that the subscript indices in the above are free. The function $f$ is then a $3\times 6$ matrix whose columns  satisfy the homogeneous Lam\'e system. 
Indeed, since after an application of the Lam\'e operator, it is possible to retrieve through the definition of $G$:
\begin{eqnarray*}
-L(\nabla_{\BW}) f(\BW)&=&\int_\Omega \frac{\partial}{\partial Z_n}(\Xi_{ap}(\nabla_{\BZ})\delta(\BW-\BZ)\BI_3) \BGT^T(\BZ)d\BZ \\
&=&\int_\Omega \delta(\BZ-\BW) \frac{\partial}{\partial Z_n}\Xi_{ap}(\nabla_{\BZ})\BGT^T(\BZ)d\BZ\;.
\end{eqnarray*}
Now, when considering the cases $\BW\in \cup_{j=1}^N \overline{B^{(j)}}$ and $\BW\in \Omega \backslash \cup_{j=1}^N \overline{B^{(j)}}$, the definition of $\BGT$ shows that the above right-hand side is equal to $\mathbb{O}_{3\times 6}$.

Again the definition of $G$ also ensures that $f(\BW)=\mathbb{O}_{3\times 6}$ for $\BW \in \partial \Omega$. An application of the Betti's formula to $f(\BW)$ and Green's matrix $G$ in $\Omega$ shows then shows that $f(\BW)=\mathbb{O}_{3\times 6}$ for $\BW \in \Omega$ and the proof of (\ref{zero_id_1}) is complete.
\hfill $\Box$

\subsection{The estimate for (\ref{SP_1})}\label{sec9.3}
Relation (\ref{QF}) is then a result of  the next Lemma:

\begin{lem}\label{lemQFest}
The relation 
\begin{equation}\label{eq59duplicate}
\langle \BM\BC, \BS\BM\BC \rangle= -\frac{h}{(8\pi)^2 (\frac{d}{4})^6 (\lambda+4\mu)^2}
\;,
\end{equation}
is valid, where
\begin{equation}\label{h2}
h=\sum_{j=1}^N  (\BM_\varepsilon^{(j)}\BC^{(j)})_p\CA_{pq}^{(j,j)} (\BM_\varepsilon^{(j)}\BC^{(j)})_q\;,
\end{equation}
with  repeated subscript indices being regarded as the indices of summation and 
\begin{equation}
\label{QFest}
|\langle \BM\BC, \BS\BM\BC \rangle|\le \text{\emph{Const }}d^{-3} \langle \BM\BC,\BM\BC \rangle\;.
\end{equation}
\end{lem}
\vspace{0.1in}\emph{Proof. Representations $(\ref{eq59duplicate})$ and $(\ref{h2})$. }The combination of (\ref{lem_DDG}) and (\ref{SP_1}) then delivers the expression
\begin{equation}\label{MC_scalar_p}
\langle \BM\BC, \BS\BM\BC \rangle= \frac{g-h}{(8\pi)^2 (\frac{d}{4})^6 (\lambda+4\mu)^2}\;,
\end{equation}
with
\begin{eqnarray}\label{h1}
g&=&\sum_{j=1}^N\sum_{k=1}^N (\BM_\varepsilon^{(j)}\BC^{(j)})_p\CA_{pq}^{(j,k)}
(\BM_\varepsilon^{(k)}\BC^{(k)})_q   
\end{eqnarray}
and $h$ is defined in (\ref{h2}).

The above expression for $g$ admits a 
form
that makes use of (\ref{Xi}) and (\ref{Xi_lin})
\begin{eqnarray}
g&=&36^2(\lambda+4\mu)^2 \CK^{(1)}+180(\lambda+4\mu)(\lambda+\mu)\CK^{(2)}-36(\lambda+4\mu)(\lambda-\mu)\CK^{(3)}\nonumber \\
&&+225(\lambda+\mu)^2 \CK^{(4)}-90(\lambda^2-\mu^2)\CK^{(5)}+9(\lambda-\mu)^2\CK^{(6)}\;,\label{h12}
\end{eqnarray}
where
\begin{eqnarray*}
&& \CK^{(1)}=\int_\Omega \int_\Omega \GF_p(\BZ)\GF_q(\BW) \Xi_{ap}(\nabla_{\BZ})\Xi_{bq}(\nabla_{\BW})G_{ab}(\BZ,\BW)\,d\BW d\BZ\;,\\
&& \CK^{(2)}=\int_\Omega \int_\Omega \Theta_{pt}(\BZ)\GF_q(\BW)\frac{\partial}{\partial Z_a}(\Xi_{ap}(\nabla_{\BZ})\Xi_{bq}(\nabla_{\BW})G_{tb}(\BZ,\BW))\,d\BW d\BZ\;,\\
&&\CK^{(3)}=\int_\Omega \int_\Omega\Theta_{ps}(\BZ)\GF_q(\BW)\frac{\partial}{\partial Z_s}(\Xi_{ap}(\nabla_{\BZ})\Xi_{bq}(\nabla_{\BW}) G_{ab}(\BZ,\BW))\,d\BW d\BZ\;,\\
&& \CK^{(4)}=\int_\Omega \int_\Omega\Theta_{pt}(\BZ)\Theta_{qs}(\BW)\frac{\partial^2}{\partial Z_a \partial W_b}(\Xi_{ap}(\nabla_{\BZ})\Xi_{bq}(\nabla_{\BW})G_{ts}(\BZ,\BW))\,d\BW d\BZ\;,\\
&& \CK^{(6)}=\int_\Omega \int_\Omega \Theta_{ps}(\BZ)\Theta_{qt}(\BW)\frac{\partial^2 }{\partial Z_s\partial W_t}(\Xi_{ap}(\nabla_{\BZ})\Xi_{bq}(\nabla_{\BW})G_{ab}(\BZ,\BW))\, d\BW d\BZ\;.
\end{eqnarray*}

Now, the term $\CK^{(2)}$ is rewritten using the Kronecker delta as
\[\CK^{(2)}=\int_\Omega  \delta_{mp}\delta_{an}\GF_q(\BW)\Xi_{bq}(\nabla_{\BW}) \int_{\Omega} \Theta_{mt}(\BZ)\frac{\partial}{\partial Z_n}(\Xi_{ap}(\nabla_{\BZ})G_{tb}(\BZ,\BW))\,d\BZ
 d\BW\;,\]
where  as shown in Lemma \ref{lem_aux_1}, the inner integral is zero. Thus $\CK^{(2)}=0$. Similar conversions and Lemma \ref{lem_aux_1} also show that the terms $\CK^{(1)}$, $\CK^{(j)}$, $ 3\le j\le 6$ are equal to zero. In this way, we have shown that $g=0$ (see (\ref{h1})).
The proof of (\ref{eq59duplicate}) and (\ref{h2}) is  complete. 

\emph{Estimate for $h$. } Next,  to prove (\ref{QF}), the estimate for the quantity $h$
\[|h|\le \text{Const }d^3 \langle \BM\BC, \BM\BC\rangle\;,\]
is proved.




To show this, an estimate for the terms $\CJ^{(m,j,j)}_{pq}$, $1\le m\le 6$ is needed that make use of the fact that for $\Bx,\By\in \Omega$, $\|G(\Bx,\By)\|=O(|\Bx-\By|^{-1})$.  Employing this, a majorant for $\CJ^{(1,j,j)}_{pq}$ is given by
\begin{equation}\label{est_Js}
\CJ^{(1,j,j)}_{pq}\le \text{Const }\int_{B^{(j)}}\int_{B^{(j)}} \frac{d\BW d\BZ}{|\BZ-\BW|^3}\le \text{Const }d^{3}\;.
\end{equation}
The estimates for $\CJ^{(m,j,j)}_{pq}$, $2\le m\le 6$ are developed in a similar way. 
Next, consider the term
\[\sum_{j=1}^N (\BM^{(j)}_\varepsilon \BC^{(j)})_p\CJ^{(m,j,j)}_{pq}(\BM^{(j)}_\varepsilon \BC^{(j)})_q, \quad, 1\le m \le 6\;.\]
Recalling that subscript indices are the indices of summation, and repeatedly applying the Cauchy inequality the above admits the inequality
\[\sum_{j=1}^N (\BM^{(j)}_\varepsilon \BC^{(j)})_p\CJ^{(m,j,j)}_{pq}(\BM^{(j)}_\varepsilon \BC^{(j)})_q \le \sum_{j=1}^N|\BM_\varepsilon^{(j)}\BC^{(j)}|^2 \left(\sum_{p,q=1}^3 (\CJ^{(m,j,j)}_{pq})^2\right)^{1/2}\;,\]
for $1\le m\le  6$. The preceding combines with (\ref{est_Js}) and the estimates for $\CJ^{(m,j,j)}_{pq}$, $2\le m\le 6$ to  show that
\[\sum_{j=1}^N (\BM^{(j)}_\varepsilon \BC^{(j)})_p\CJ^{(m,j,j)}_{pq}(\BM^{(j)}_\varepsilon \BC^{(j)})_q \le \text{Const } d^{3}\sum_{j=1}^N|\BM_\varepsilon^{(j)}\BC^{(j)}|^2 \;.\]
Therefore, consulting (\ref{eq59duplicate}) it can be asserted that (\ref{QFest}) holds. Thus the proof of the present lemma and (\ref{QF}) is complete. 
\hfill $\Box$

\section*{Appendix C: Explicit representation of dipole fields  for spherical cavities} 
\label{secQsphere}
\renewcommand{\theequation}{C.\arabic{equation}}
\setcounter{equation}{0}

It is shown in this section that for certain geometries, model fields used in the asymptotic approximations presented here can be constructed in the closed form.
Here, it is assumed that the voids $\omega_\varepsilon^{(j)}$, $j=1,\dots, N,$ are spherical cavities.
The matrix $\BQ^{(k)}_\varepsilon$ for a spherical cavity $\omega^{(k)}_\varepsilon$, with radius $a^{(k)}_\varepsilon$ and centre at $\BO^{(k)}=\{O^{(k)}_i\}_{i=1}^3$, in an infinite solid can be reconstructed using the approach presented in \cite{Lure} that makes use of Papkovich-Neuber potential representation   for solutions to three-dimensional elasticity problems. 

In this case, the matrix takes the form
\begin{eqnarray}
\BQ^{(k)}_\varepsilon(\Bx)&=&-(\BGX(\nabla_\Bx)^T \Gamma(\Bx,\BO^{(k)}))^T\BM^{(k)}_\varepsilon+\frac{1}{|\Bx-\BO^{(k)}|^5}\BGX(\Bx-\BO^{(k)})\mathfrak{A}^{(k)}_1+\frac{1}{|\Bx-\BO^{(k)}|^7}\Big\{\BGX(\Bx-\BO^{(k)})\mathfrak{Y}(\Bx-\BO^{(k)})\nonumber\\
&&+{(x_1-O^{(k)}_1)(x_2-O^{(k)}_2)(x_3-O^{(k)}_3)}\mathfrak{A}^{(k)}_2+\mathfrak{M}(\Bx-\BO^{(k)})\BGX(\Bx-\BO^{(k)})\mathfrak{A}^{(k)}_3\Big\}\;.\nonumber\\
\label{Qk_sphere_1}
\end{eqnarray}
Here the dipole matrix $\BM^{(k)}_\varepsilon$ is given in (\ref{Meps1}), (\ref{Meps2}), and  the matrices $\mathfrak{A}^{(k)}_p$, $1\le p\le 3$ are
\[\mathfrak{A}^{(k)}_1=-\frac{3(\lambda+\mu)(a^{(k)}_\varepsilon)^5}{9\lambda+14\mu}\left[\begin{array}{ccccc}
\mathfrak{B}^{(1)}&&      && \mathbb{O}_{3} \\
 \mathbb{O}_{3}&&      && {2}\,\mathbb{I}_{3}\\
\end{array}\right]\;, \qquad \mathfrak{B}^{(1)}=\left[\begin{array}{ccccc}
3&&1 && 1\\
1&& 3&&1 \\
1&&1 &&3 \end{array}
\right]\;,\]
\[ \mathfrak{A}^{(k)}_2=\frac{15\sqrt{2}(\lambda+\mu)(a^{(k)}_\varepsilon)^5}{9\lambda+14\mu}\left[\begin{array}{ccccc}
 \mathbb{O}_{3}&&      && \mathfrak{B}^{(2)}\\
\end{array}\right]\;, \qquad \mathfrak{B}^{(2)}=\left[\begin{array}{ccccc}
0&&0&& 1\\
0&& 1&&0 \\
1&&0 &&0
 \end{array}
\right]\;,\]
\[ \mathfrak{A}^{(k)}_3=\frac{30(\lambda+\mu)(a^{(k)}_\varepsilon)^5}{9\lambda+14\mu}\left[\begin{array}{ccccc}
\mathbb{O}_3&&      && \mathbb{O}_{3} \\
 \mathbb{O}_{3}&&      && \mathbb{I}_{3}\\
\end{array}\right]\;. \]
Also the matrix functions in (\ref{Qk_sphere_1}), $\mathfrak{Y}$ and $\mathfrak{M}$, are given as
\[\mathfrak{M}(\Bx)=\left[\begin{array}{ccccc}
x_1^2&& 0 &&0\\ 
0&&x_2^2&&0\\ 
0&&0&&x_3^2
\end{array}\right]\;, \]
and 
\[\mathfrak{Y}(\Bx)=\frac{15(\lambda+\mu)(a^{(k)}_\varepsilon)^5}{9\lambda+14\mu}\left[
\begin{array}{ccc}
\mathfrak{D}(\Bx)&& \mathbb{O}_{3}\\
 \mathbb{O}_{3}&& \mathbb{O}_{3}
\end{array}\right]\;, \qquad \mathfrak{D}(\Bx)=\left[\begin{array}{ccccc}
x_{ 1}^2&& x_2^2&&x_3^2  \\
x_1^2&&x_2^2&&x_3^2\\ 
x_1^2&&x_2^2&&x_3^2
\end{array}\right]=\left[\begin{array}{ccccc}
1&& 1&&1\\
1&&1&&1\\
1&&1&&1
\end{array}\right]\mathfrak{M}(\Bx)\;.\]

\end{document}

%% file: macroarx.tex
\newcommand{\bfm}[1]{\mbox{\boldmath ${#1}$}}

\newcommand{\beqa}{\begin{eqnarray}}
\newcommand{\eeqa}{\end{eqnarray}}
\newcommand{\bequ}{\begin{equation}}
\newcommand{\eequ}[1]{\label{#1}\end{equation}}

\newcommand{\Gve}{\varepsilon}

\newcommand{\GF}{\Phi}

\newcommand{\GO}{\Omega}

\newcommand{\BGf}{\bfm\phi}

\newcommand{\BGn}{\bfm\eta}

\newcommand{\BGs}{\bfm\sigma}

\newcommand{\BGx}{\bfm\xi}

\newcommand{\BGy}{\bfm\psi}

\newcommand{\BGF}{\bfm\Phi}

\newcommand{\BGT}{\bfm\Theta}

\newcommand{\BGX}{\bfm\Xi}
\newcommand{\BGY}{\bfm\Psi}

\newcommand{\CA}{{\cal A}}

\newcommand{\CJ}{{\cal J}}
\newcommand{\CK}{{\cal K}}
\newcommand{\CL}{{\cal L}}
\newcommand{\CM}{{\cal M}}
\newcommand{\CN}{{\cal N}}

\newcommand{\CV}{{\cal V}}

\newcommand{\BCM}{{\bfm{\cal M}}}

\def\Be{{\bf e}}
\def\Bf{{\bf f}}

\def\Bn{{\bf n}}

\def\Br{{\bf r}}

\def\Bu{{\bf u}}
\def\Bv{{\bf v}}
\def\Bw{{\bf w}}
\def\Bx{{\bf x}}
\def\By{{\bf y}}
\def\Bz{{\bf z}}
\def\BA{{\bf A}}
\def\BB{{\bf B}}
\def\BC{{\bf C}}
\def\BD{{\bf D}}
\def\BE{{\bf E}}
\def\BF{{\bf F}}

\def\BI{{\bf I}}
\def\BJ{{\bf J}}

\def\BM{{\bf M}}
\def\BN{{\bf N}}
\def\BO{{\bf O}}
\def\BP{{\bf P}}
\def\BQ{{\bf Q}}
\def\BR{{\bf R}}
\def\BS{{\bf S}}

\def\BU{{\bf U}}
\def\BV{{\bf V}}
\def\BW{{\bf W}}

\def\BZ{{\bf Z}}

\newcommand{\beq}{\begin{equation}}
\newcommand{\eeq}{\end{equation}}
\newcommand{\overliner}{\begin{eqnarray}}
\newcommand{\earr}{\end{eqnarray}}
\newcommand{\beqn}{\begin{equation*}}
\newcommand{\eeqn}{\end{equation*}}
\newcommand{\overlinern}{\begin{eqnarray*}}
\newcommand{\earrn}{\end{eqnarray*}}
\newcommand{\prt}{\partial}

%% file: mesoscale_voids_text_arXiv.bbl
\begin{thebibliography}{88}
\bibitem{ADNI} S. Agmon, A. Douglis, L. Nirenburg, ``Estimate near the boundary for solutions of elliptic partial differential equations satisfying general boundary conditions. I.,'' Communications on Pure and Applied Mathematics, vol. xii, 623--727, 1959.
\bibitem{ADNII} S. Agmon, A. Douglis, L. Nirenburg, ``Estimate near the boundary for solutions of elliptic partial differential equations satisfying general boundary conditions. II.,'' Communications on Pure and Applied Mathematics, vol. xvii, 35--92, 1964.

\bibitem{Bakhvalov}  
N. S. Bakhvalov and G. Panasenko,
\textit{Homogenization: Averaging Processes in Periodic Media},
Springer, Berlin (1989).

\bibitem{BramPayne}  
J. H. Bramble and
L. E.  Payne,
``Some mean value theorems in elastostatics,''
\textit{J. Soc.
Indust. Appl. Math.}
\textbf{12}, No.  1, 105--114
 (1953).


\bibitem{Ciarlet}
P.G. Ciarlet, ``Mathematical elasticity. Volume II: theory of plates,'' North-Holland, Amsterdam, (1997)

\bibitem{Ciarlet_Destuyunder}
 P.G. Ciarlet and P. Destuynder, ``A justification of the two-dimensional plate model,'' J. M\'ec, \textbf{18}, 315--344, (1979).
 
 

\bibitem{Fig_1} 
R. Figari, E. Orlandi, and A.  Teta,
``The Laplacian in regions with
many small obstacles: fluctuations around the limit operator,''
\textit{J. Statist. Phys.}
\textbf{41}, No.  3-4, 465--487
 (1985).

\bibitem{Fig_2} 
R. Figari and A.  Teta,
``A boundary value problem of mixed type on perforated domains,''
\textit{Asymptotic Anal.}
\textbf{6}, No.  3, 271--284
 (1993).

\bibitem{Horgan} C.O. Horgan and L.E. Payne: ``On inequalities of Korn, Friedrichs and Babu\v ska-Aziz,'' \emph{Archive for Rational Mechanics and Analysis} \textbf{11} no. IV, issue 2, 165--179, (1983)
\bibitem{MarKhrus}  
V. A. Marchenko and  E. Y. Khruslov,
\textit{Homogenization of Partial Differential Equations}, Birkh\"auser,
Basel
 (2006).
 
\bibitem{KMMI} V.A. Kozlov, V.G. Maz'ya,  A.B. Movchan: ``Asymptotic representation of elastic fields in a multi-structure,'' \emph{Asymptotic Anal.}, {\bf 11}, 343--415, (1995).
\bibitem{KMMII} V.A. Kozlov, V.G. Maz'ya,  A.B. Movchan: ``Asymptotic Analysis of Fields in Multi-Structures,'' Oxford Mathematical Monographs,  Clarendon Press, Oxford (1999).
 
 \bibitem{Lure} A.I. Lur'e, ``Three-Dimensional Problems of the Theory of Elasticity,''  Interscience, New York, (1964).  
  \bibitem{CRM}  
V. Maz'ya and A.  Movchan,
``Uniform asymptotic formula for regularly
and singularly perturbed domains,''
\textit{C. R. Math. Acad. Sci. Paris}
\textbf{343}, No.  3, 185--190 (2006).

\bibitem{JCOM}  
V. Maz'ya and A.  Movchan,
``Uniform asymptotic formulae for Green's
functions in singularly perturbed domains,''
\textit{J. Comput. Appl. Math.}
\textbf{208}, No. 1, 194--206 (2007).

\bibitem{MMAS}  
V. Maz'ya and A.  Movchan,
``Uniform asymptotic approximations of
Green's function in a long rod,''
\textit{Math. Methods Appl. Sci.}
\textbf{31}, No.  17, 2055--2068 (2008).

\bibitem{Sob_vol} 
V. Maz'ya and A.  Movchan,
``Uniform asymptotics of Green's kernels for
mixed and Neumann problems in domains with
small holes and inclusions. Sobolev spaces in mathematics. III,''
pp.  277-213,
In:
\textit{Sobolev Spaces
in Mathematics. III. Applications in Mathematical Physics},
Springer and Tamara Rozhkvoskaya Publisher,  New York etc
(2009).


\bibitem{Maz_MN} 
V. Maz'ya and A.  Movchan,
``Asymptotic treatment of perforated domains without
homogenization,''
\textit{Math. Nachr.}
\textbf{283}, No.  1, 104--125 (2010).



\bibitem{RAN} 
V. Maz'ya, A.   Movchan,  and  M.  Nieves,
``Uniform asymptotic formulae
for Green's tensors in elastic singularly perturbed
domains with multiple inclusions,''
\textit{Rend. Accad. Naz. Sci. XL Mem. Mat. Appl.}
\textbf{5}, No.  30, 103--157
 (2006).

\bibitem{AA} 
V. G.  Maz'ya, A. B.  Movchan,    and  M. J. Nieves,
``Uniform asymptotic formulae for Green's
tensors in elastic singularly perturbed domains,''
\textit{Asymptot. Anal.}
\textbf{52},
No.
3-4, 173--206, (2007).

\bibitem{AMS_tran} 
V. Maz'ya, A.   Movchan,  and  M.  Nieves,
``Green's kernels for transmission problems in
bodies with small inclusions,''
In:
\textit{Operator Theory and Its Applications},
pp.  127--160,
Am. Math. Soc.,
Providence, Ri
 (2010).

\bibitem{Maz_MMS} 
V. Maz'ya, A.   Movchan,  and  M.  Nieves,
``Mesoscale asymptotic approximations to solutions
of mixed boundary values problems in perforated domains,''
\textit{Multiscale Model. Simul.}
\textbf{9}, No.  1, 424--448
 (2011).

\bibitem{MMN_book} V. Maz'ya, A. Movchan and M. Nieves: \textit{Green's Kernels and Meso-Scale Approximations in Perforated Domains}, Lecture Notes in Mathematics \textbf{2077}, Springer, 2013

\bibitem{MMN_Mesoelast} 
V. Maz'ya, A.   Movchan,  and  M.  Nieves,
``Mesoscale approximations for solutions
of the Dirichlet problem in a perforated elastic body,'' 
\textit{Journal of Mathematical Sciences},\textbf{202}, no.2, 215--254, (2014).



\bibitem{MMP} A.B. Movchan, N.V. Movchan, C.G. Poulton: \textit{Asymptotic Models of Fields in Dilute and Densely Packed Composites}, Imperial College Press, (2002).


 \bibitem{OPTH1} 
V. Maz'ya,  S.  Nazarov,  and B.  Plamenevskij,
\textit{Asymptotic Theory of Elliptic Boundary
Value Problems in Singularly Perturbed Domains. I},
Birkh\"auser,  Basel (2000).

\bibitem{OPTH2}  
V. Maz'ya,  S.  Nazarov,  and B.  Plamenevskij,
\textit{Asymptotic Theory of Elliptic Boundary
Value Problems in Singularly Perturbed Domains. II},
Birkh\"auser,  Basel (2000).


\bibitem{Ozawa} S. Ozawa,
``Approximation of Green's function in a region with many obstacles,'' \textit{Geometry and Analysis on Manifolds}, Ed: T. Sunada, Lecture Notes in Mathematics, \textbf{1339}, (Springer, New York, 1988), 212--225.

\bibitem{SP} E. S\'anchez-Palencia, ``Homogenisation Method for the study of composite Media. Asymptotic Analysis II,'' \textit{Lecture Notes in Math.}, \textbf{985}, (Springer, Berlin, 1983), 192--214.

\bibitem{Stein} E.M. Stein, ``\emph{Singular Integrals and Differentiability Properties of Functions},'' Princeton University Press, New Jersey, 1970.
\end{thebibliography}
